\documentclass[pre,twocolumn]{revtex4}

\usepackage{epsfig,amsmath,amssymb,graphicx,color,calc}
\usepackage{axodraw2}     
\usepackage{pstricks}         
\usepackage{color}             

\newcommand{\beq}{\begin{equation}}
\newcommand{\eeq}{\end{equation}}

\let\a=\alpha \let\b=\beta    \let\d=\delta 
\let\e=\varepsilon         
 \let\m=\mu  \let\n=\nu       \let\p=\pi  
\let\s=\sigma         
        \let\G=\Gamma

 \let\r=\rho \let\th=\theta 

\let\io=\infty
 \def\OO{{\cal O}} \def\MM{{\cal{M}}}
\def\FF{{\cal F}}
\def\to{\rightarrow}
\def\la{\left\langle}
\def\ra{\right\rangle}
\def\Tr{{\rm{Tr}~}}
\def\bfr{{\boldsymbol{\rho}}}
\def\bfn{{\boldsymbol{\nu}}}

\def\bfh{{\boldsymbol{h}}}
\def\hr{{\hat{\r}}}
\def\th{{\tilde{h}}}
\def\tc{{\tilde{c}}}
\def\hro2{{{\hat{\r}}^{(2)}}}
\def\ro2{{{\r}^{(2)}}}

\begin{document}

\title{Systematic expansion in the order parameter \\ for replica theory of the dynamical glass transition}

\author{Hugo Jacquin}
\affiliation{Laboratoire Mati\`ere et Syst\`emes Complexes, UMR CNRS 7057,
Universit\'e Paris Diderot -- Paris 7, 10 rue Alice Domon et L\'eonie 
Duquet, 75205 Paris cedex 13, France}

\author{Francesco Zamponi}
\affiliation{Laboratoire de Physique Th\'eorique, Ecole Normale Sup\'erieure, 
UMR CNRS 8549, 24 Rue Lhomond, 75231 Paris Cedex 05, France}

\date{\today}

\begin{abstract}
It has been shown recently that predictions from Mode-Coupling Theory for the glass transition of hard-spheres become 
increasingly bad when dimensionality increases, whereas replica theory predicts a correct scaling. Nevertheless if one focuses on the regime 
around the dynamical transition in three
dimensions, Mode-Coupling results are far more convincing than replica theory predictions. It seems thus necessary to reconcile the 
two theoretic approaches in order to obtain a theory that interpolates between low-dimensional, Mode-Coupling results, 
and ``mean-field" results from replica theory. Even though quantitative results for the dynamical transition issued from
replica theory are not accurate in low dimensions, two different approximation schemes --small cage expansion and replicated 
Hyper-Netted-Chain (RHNC)-- provide the correct 
qualitative picture for the transition, namely a discontinuous jump of a static order parameter from zero to a 
finite value. The purpose of this work is to develop a systematic expansion around the RHNC result 
in powers of the static order parameter, and to calculate the first correction in this expansion. Interestingly, this correction involves the static three-body correlations of the liquid. More importantly, we separately demonstrate that higher order terms in the expansion are quantitatively relevant at the transition, and that the usual mode-coupling kernel, involving two-body direct correlation functions of the liquid, cannot be recovered from static computations.
\end{abstract}



\maketitle

\tableofcontents

\section{Introduction}
\label{intro}

A lot of theoretical activity has been devoted to the general problem of the glass transition in the last decades~\cite{BB11}. Amongst all theories, only two of them emerge 
from a microscopic description, and without assuming the existence of the glass transition itself. The first theory is an adaptation of the mode-coupling 
theory of critical dynamics~\cite{HH77} to the glass transition, and was devised by G\"otze and collaborators~\cite{GS92}. It is nowadays simply denoted by Mode-Coupling 
Theory (MCT). The other one is the Random-First-Order Transition theory (RFOT)~\cite{bookRFOT}, an adaptation to molecular glasses of the mean-field replica theory of 
spin glasses~\cite{BY86,MPV87}. While the development of a first principle theory of dynamics 
is plagued with difficulties
(in particular for MCT, this is mainly due to its lack of self-consistency)~\cite{RC05,CR06,ABL06,MCTclothes,JW11}, 
replica theory is a static theory, hence calculations are easier, and it has been adapted to the study of the dynamical arrest of structural glasses, 
{\it via} three different approximation schemes: the small cage expansion 
\cite{MP99b,MP99a}, the effective potential approximation~\cite{PZ05,PZ10} and the replicated Hyper-Netted-Chain (RHNC) approximation~\cite{MP96,CFP96}. 

The first two
are formulated in terms of the local cage size around a particle. This parameter is expected to be infinite in the liquid phase (the particles are free to visit all space) but finite 
and small in the glass phase, thus can serve as an order parameter for the glass transition. However the large jump of this scalar order parameter (from infinity to small!) 
seems to prevent one from obtaining good quantitative results around the dynamical glass transition, whereas predictions are more robust deep inside the glass phase~\cite{MP99a,MP99b}. The situation is reversed for the RHNC approximation: the theory is formulated in terms 
of a two-point order parameter (a pair correlation function between two replicas, equivalent to the non-ergodicity parameter), that displays a jump from zero to a non-zero 
(but non small) value at the dynamical transition. The RHNC approximation correctly predicts, in a qualitative way, the 
phenomenology of the dynamical glass transition, i.e. the freezing of density fluctuations due to the appearance of an exponential number of 
metastable states in the system~\cite{MP96,CFP96}. 
However, the neglect of some important corrections (see~\cite{PZ10} for a detailed discussion) 
both prevents to obtain accurate quantitative results near the glass transition, 
and gives inconsistent results deep inside the glass phase~\cite{MP96}. The main problem is that the structural 
properties of the glass (and in particular the non-ergodicity parameter) are found to be quantitatively incorrect. For example, the non-ergodicity parameter for hard spheres 
shown in~\cite{PZ10} and in Fig.~\ref{fig:fk_RHNC} of this paper poorly compares to the experimental ones found for example in~\cite{MUP91}.

The difficulties of the replica method in treating the dynamical glass transition, compared to the quantitative success of MCT in the same regime~\cite{Go99}, must not hide 
its successes: it is able, within a purely static framework, and starting from the Hamiltonian of the system, to predict without assuming it the appearance of a large number of 
metastable states, to quantify this number through the complexity, and to make predictions for the critical properties of the long-time dynamics 
\cite{CFLPRR12,FJPUZ12a,FJPUZ12b}. It is thus of primary importance to continue developing the RHNC approximation scheme in order to probe the limits of this 
replica (static) approach, and in order to set the stage for an hypothetical theory that would be able to reconcile dynamic, MCT-like computations, and static, replica 
computations. Interestingly, a first bridge was made between the two~\cite{Sz10} by looking at the dynamics in a replicated liquid, and treating these dynamics in a MCT 
fashion. Unfortunately, this approach suffers from the same lack of internal self-consistency as the original MCT computation, and it only succeeds in reinforcing the general
belief that MCT and replica theory can indeed be unified in a coherent scheme (still neglecting activated events), as was done in the context of spin glasses~\cite{KW87}.

In this paper we perform a first step in the general task of making RHNC an efficient quantitative theory of the dynamical transition, by setting up a perturbative expansion in 
powers of the static order parameter used in replica theory to characterize the glass phase. This order parameter is proportional to the non-ergodicity factor, and our 
expansion thus corresponds to the ``weak glass" expansion sought for in~\cite{BB09}. We calculate the first correction to RHNC in this expansion, and show that it already leads 
to some improvement of the quantitative results, while the qualitative picture remains stable. Interestingly, we show that this new term, and the subsequent ones, involve
the three-body and higher order correlations of the liquid, which currently receive renewed interest: even though two-body static correlations of the liquid are blind to the 
presence of the glass transition, it is possible that three-body functions are more sensitive to it~\cite{Co11}. Our work paves the way to more powerful 
re-summations, maybe allowing to unify the three versions of replica theory for structural glasses (small cage, effective potential, and RHNC) in a unique scheme, and 
possibly obtain a quantitatively competitive theory of the dynamical transition.

The paper is organized as follows.
In Section~\ref{sec2} we briefly review the replica interpretation of the glass transition, focusing on the two main order parameters, static and dynamic, 
currently used to describe it in a theoretical way. In Section~\ref{sec3} we introduce the replicated liquid theory that will allow us to perform 
approximations on the static order parameter, and briefly review its simplest implementation, the replicated Hyper-Netted-Chain approximation~\cite{MP96}. 
We then explain how to proceed from this approximation to perform an expansion in powers of the static order parameter. In Section~\ref{sec4}, the
first correction to the replicated Hyper-Netted-Chain approximation is obtained, and shown to involve three-body correlation functions of the liquid. 
This section is mostly technical and can be safely jumped by the reader uninterested in details of the computation. A numerical solving of this 
improved Hyper-Netted-Chain approximation is presented in Section~\ref{sec5}, and we conclude in Section~\ref{conclusion}.

\section{Static order parameter for glasses}
\label{sec2}

The glass transition is {\it a priori} defined in a dynamical way as the temperature below which (or density above which) density fluctuations become
frozen in the liquid. However, replica computations on mean-field spin glasses have given an insight into the 
thermodynamics of glasses, and provided us with a static order parameter for the glass transition~\cite{MPV87}. 
In this section we recall the procedure needed
to define and calculate this order parameter, and show its relation with the dynamical order parameter traditionally used in glass physics, namely the 
non-ergodicity parameter.

\subsection{Order parameter with replicas}

In order to detect the glass transition from a static observable, one introduces identical copies of the system of interest.
Consider an integer number $m$ of copies of the system, i.e. a liquid of $m \times N$ particles. The copies are indexed by alphabetical 
characters $a,b,\ldots$, and the position of particle $i$ in copy $a$ will be denoted by $x_i^{a}$. Within each copy, all the 
particles interact with a given pair potential, but each particle in a given copy is 
attracted to all other particles of all other copies via an attractive pair potential of infinitesimal amplitude $\e$. Consider for example copy $a$ and 
copy $b$, with $a \ne b$. In analogy with usual liquid theory (see~\cite{hansen} and Eq.~(\ref{def_micro_density}) in the following), each of them has a microscopic density 
defined by:
\beq
\hr_a(x) = \sum_{i=1}^N \d(x-x_i^a) \ , \label{def_roa}
\eeq
Consider now the generalization of the pair correlation function, as defined in~\cite{hansen} and Eq.~(\ref{def_h}) below, to a replicated system:
\beq
\left\{ \begin{array}{ll}
h_{ab}(x,y) & = \displaystyle \frac{\la \hr_a(x) \hr_b(y) \ra}{\r^2} - 1 \text{ for } a \ne b \ , \\
& \\
h_{aa}(x,y) & = \displaystyle \frac{\la \hr_a(x) \hr_a(y) \ra}{\r^2} - 1 - \frac 1 \r \d(x-y) \ .
\end{array} \right. \label{def_hab}
\eeq
Note that, for $a \ne b$, the coincident point term is absent, because particle $i$ of copy $a$ interacts with all particles of
copy $b$, including particle $i$, whereas within copy $a$, particle $i$ does not interact with itself. 

We want to study what happens in the thermodynamic limit $N\to\io$ when the coupling potential strength $\e$ is vanishingly small.
As usual, a phase transition will be signaled by the fact that the limits $N\to\io$ and $\e\to 0$ do not commute~\cite{KT89,Mo95,MP00}.
More precisely, if we suppose that a dynamical transition occurs at $\r_d$, with a mean-field phenomenology, we will
have that:
\begin{itemize}
\item if $\r < \r_d$ and letting $\e$ go to zero, the two copies de-correlate, i.e. $h_{ab} = 0$, and the system is not trapped in a metastable state.
In this case the limit $\e\to 0$ and $N\to\io$ commute.
\item if $\r > \r_d$, the two copies are trapped into the same metastable state by their mutual attraction $\e$. Letting $\e$ go to zero after $N\to\io$, 
the two copies will stay correlated and $h_{ab} \ne 0$. Note that obviously the copies will de-correlate if $\e$ is set to zero before $N\to\io$, hence
the two limits do not commute in the glass phase.
\end{itemize}
Of course this transition is observed only if metastable states are present in the system, that are able to constrain copies to stay
in the same state, in the thermodynamic limit and in the long-time limit. As we will see in the following, this assumption can be 
checked self-consistently within the theory.
In order to recover the equilibrium properties of the original, non-replicated system, one has to make an analytic continuation 
to non-integer values of $m$, the number of replicas, and take the limit $m \to 1$ at the end of the calculations~\cite{Mo95,MP96}.

The general procedure is then the following:
\begin{itemize}
\item Compute the equilibrium properties of a $m$-times replicated liquid, with small attraction $\e$ between different copies
\item Take the thermodynamic limit, and compute the desired quantities as functions of $m$
\item Send the attraction $\e$ to zero
\item Take the limit $m \to 1$ to recover the original system
\end{itemize}

We show in the following that calculating the static order parameter defined above with this prescription amounts to evaluating the non-ergodicity parameter
of the glass.

\subsection{Link with the dynamic order parameter}

In the mean-field replica interpretation, the partition function of the system is supposed to be separated into many pure states, that correspond 
to minima of the free-energy. Because of this separation, averages can be separated in two operations:
first an average inside a state $\a$, denoted by $\la \bullet \ra_\a$, and then an average over all states denoted \mbox{by 
$\overline{~ \bullet~ }$.} Calculating $h_{ab}(x,y)_\e$, the correlation between copy $a$ and $b$ (with $a \ne b$), we take into account that
the attractive coupling between $a$ and $b$ will force them into the same state $\a$ leading to:
\beq
h_{ab}(x,y)_{\e} = \frac{\overline{\la \hr(x) \hr(y) \ra_{\a,\e}}}{\r^2} - 1 \ .
\eeq
Letting the interaction go to zero will allow the two replicas to de-correlate inside the state, leading to:
\beq
\lim_{\e \to 0} h_{ab}(x,y)_{\e} =  \frac{\overline{\la \hr(x) \ra_\a \la \hr(y) \ra_\a}}{\r^2} - 1 \ .
\eeq
Note that the average density inside a state need not be constant, because translational invariance is restored only after
summation over all the states. 

Independently, the time-dependent density-density correlation that is the focus of MCT, and more generally dynamical calculations, 
is defined by:
\beq
F(x,y,t) = \la (\hr(x,t) - \r) (\hr(y,0) - \r) \ra \ ,
\eeq
where at $t=0$ the system is taken at equilibrium.
Note that the initial time value of this function is related to the inverse Fourier transform of the structure factor:
\beq
F(x,y,0) = \r^2 h(x,y) + \d(x-y) \r \ .
\label{def_Fk0}
\eeq
Above the glass transition, and in the long time limit, the system eventually gets stuck in a metastable state $\a$, and we 
have:
\beq
F(x,y,t) \underset{t \to \io}{\to} \overline{\la \hr(x) \hr(y) \ra_{\a}} - \r^2 \ .
\eeq
But the system is at least able to de-correlate inside the state, giving:
\beq
F(x,y,t) \underset{t \to \io}{\to} \overline{ \la \hr(x) \ra_{\a} \la \hr(y) \ra_{\a} } - \r^2 = \r^2 h_{ab}(x,y) \ .
\label{link_Fkt_htilde}
\eeq
Now taking the Fourier transform of equations~(\ref{def_Fk0}) and~(\ref{link_Fkt_htilde}) we get:
\beq\begin{split}
& F(k,t) \underset{t \to \io}{\to} \r^2 h_{ab}(k) \ , \\
& F(k,0) = \r S(k) \ , 
\end{split}\eeq
where $S(k)$ is the structure factor defined in~\cite{hansen} and Eq.~(\ref{def_Sk}) below.
Finally the non-ergodicity factor is traditionally defined as:
\beq
f(k) \equiv \lim_{t \to \io} \frac{F(k,t)}{F(k,0)} \ ,
\eeq
so that we get~\cite{PZ10,Sz10}:
\beq
f(k) = \frac{\r h_{ab}(k)}{S(k)} \text{ with } a \ne b \ .
\label{link_f_hab}
\eeq
We thus find that, in the replica interpretation, the dynamical and static 
order parameter are the same physical observable. This parallel is exploited in order to compute long-time properties of the dynamics from 
replica calculations, for example in~\cite{CFLPRR12} or~\cite{FJPUZ12a,FJPUZ12b}.

\section{Expansion in powers of the order parameter}
\label{sec3}

Before trying to obtain a good theory for our replicated system and studying the glass phase, we must introduce the liquid theory tools 
used to describe the liquid phase, in order to fix notations and for the sake of completeness.

In this paper, we will consider for concreteness a system composed of $N$ spheres of diameter $D$ interacting via a hard-sphere pair potential $v$:
\beq \begin{split}
v(x,y) = \left\{ \begin{array}{ll}
\io & \text{ if } | x- y | \le D \ ,  \\
0 & \text{ otherwise} \ .
\end{array} \right.
\end{split} \eeq 
Note that we use here the hard-sphere potential mainly for the sake of simplicity, because it has only one parameter, the density, and because 
its liquid properties are well studied, both numerically and theoretically. However the general computation scheme we develop here is not restricted
to this potential and can be applied to any pairwise additive potential.

A microscopic configuration $\s$ of the system specifies all the positions $x_i$ of the particles, with $i \in [1,N]$. The 
Hamiltonian of the system is \mbox{then $H(\s) = \sum_{i<j} v(x_i-x_j)$.}
For any inverse temperature $\b = 1/k_B T$, the Boltzmann weight of such a configuration is $e^{- \b H(\s)}$, which can 
be either $0$ if there is any overlap between spheres, or $1$ if no spheres overlap. The temperature thus plays no role 
here. To stay in contact with standard liquid theory notations, we will nonetheless keep it explicit in this subsection. 
We fix units so that $D = 1$ and $k_B=1$.  

For a given chemical potential $\m$ and an eventual external potential $\Psi$, the grand canonical partition function of the system is 
defined as:
\beq
Z_{\rm{liq}} = \Tr e^{-\b H(\{x_i\}) + \sum_{i} \n(x_i)} \ ,
\eeq
where the grand-canonical trace is defined 
\mbox{as $\Tr \bullet = \sum_{N=0}^\io \frac 1 {N!} \int \prod_{i=1}^N dx_i ~ \bullet$,} 
and the generalized chemical potential is defined as $\n(x) = \b \m - \b \Psi(x)$.
The grand-canonical average of an observable $\OO$ is defined by:
\beq
\la \OO(\{x_i\}) \ra_{\rm{liq}} \equiv \frac 1 Z \Tr \OO(\{x_i\}) e^{-\b H(\{x_i\}) + \sum_{i} \n(x_i)} \ , 
\eeq
We define the microscopic density $\hr$ and the average density $\r$ as:
\beq \begin{split}
& \hr(x) = \sum_{i=1}^N \d(x-x_i) \ , \\
& \r(x) = \la \hr(x) \ra_{\rm{liq}} \ .
\label{def_micro_density} 
\end{split} \eeq 
We limit ourselves to translationally invariant systems, so that the average density is constant in the system and 
$\r(x) = \r$ in all points. The pair correlation $h$ is linked to the normalized second cumulant of the microscopic density, apart from a 
coincident point term:
\beq \begin{split}
h(x,y) & = \frac 1{\r^2} \la \sum_{i=1}^N \sum_{j \ne i} \d(x-x_i) \d(y-x_j) \ra_{\rm{liq}} - 1  \\
& = \frac{\la \hr(x) \hr(y)\ra_{\rm{liq}}}{\r^2} - 1 - \frac 1 \r \d(x-y) \ .
\label{def_h}
\end{split} \eeq 
Subtracting $1$ gives a function that decays to zero at large interparticle distance, and thus allows to compute its Fourier
transform. In a translational invariant system, $h(x,y)$ depends only on $|x-y|$, and we define the structure factor $S$, related to its Fourier 
transform as:
\beq
S(k) = 1 + \r h(k) \ . 
\label{def_Sk}
\eeq
$h$ is only the second function of a whole hierarchy of cumulants of the microscopic density. In the following we will need 
the next cumulant in this hierarchy:
\beq \begin{split}
W^{(3)}_{\rm{liq}}(x,y,z) = & \la \hr(x) \hr(y) \hr(z) \ra + 2 \r(x) \r(y) \r(z) \\
& - \r(x) \la \hr(y) \hr(z) \ra - \{ 2 ~ \rm{permutations}\} \ . 
\label{W3_liquid}
\end{split} \eeq
Usual liquid theory approximations start form the Legendre transform of $\ln Z$ with respect to the chemical potential, 
using the fact that the chemical potential is coupled to the microscopic density. Thus we define:
\beq 
\left\{ \begin{array}{ll}
& \displaystyle \G_{\rm{liq}}[\r] = \int_x \r(x) \n^*[\r](x) - \ln Z_{\rm{liq}}[\n^*[\r]] \ , \\
& \displaystyle \n_a^*[\r] \text{ such that } \left. \frac{\delta \ln Z_{\rm{liq}}[\n]}{\delta \n_a(x)} \right|_{\n^*[\r]} = \r(x) \ .
\end{array} \right.
\eeq
From the virial expansion~\cite{mayer}, one can show that this functional is composed of an ideal gas term $\G_{\rm{id}}$ plus an
excess term:
\beq \begin{split}
\G_{\rm{liq}}[\r] & \equiv \G_{\rm{id}}[\r] + \G_{\rm{ex}}[\r] \ , \\
\G_{\rm{id}}[\r] & \equiv \int_x \r(x) \left[ \ln \r(x) - 1 \right] \ .
\end{split} \eeq
From this functional, one can define a hierarchy of correlation functions (equivalent to the vertex functionals in field theory)
called the direct correlation functions~\cite{hansen}:
\beq
c^{(n)}_{\rm{liq}}(x_1,\cdots,x_N) = \frac{\delta^n \left( \G_{\rm{liq}}[\r] - \G_{\rm{id}}[\r] \right)}{\delta \r(x_1) 
\cdots \delta \r(x_N)} \ .
\label{def_c}
\eeq
By the properties of the Legendre transform (the calculation is shown formally in Eq.~(\ref{W2_G2_formal}) and can be 
straightforwardly applied to the case of liquid theory), 
the second order direct correlation function $c^{(2)}_{\rm{liq}}$, that will be denoted by 
$c$ in the following, is the inverse function of $h$, which is expressed by the Ornstein-Zernike equation:
\beq
(1 + \r h(k))(1 - \r c(k) ) = 1 \label{OZ_2} \ ,
\eeq
where $c(k)$ is the Fourier transform of $c^{(2)}_{\rm liq}(r)$ as defined in Eq.~(\ref{def_c}).
We can rewrite this equation in a more usual field-theoretic notation:
\beq
\int_{z} G_{liq}(x,z) \G^{(2)}_{liq}(z,y) = \d(x,y) \ ,
\label{OZ2_fieldtheory}
\eeq
where we have defined the propagator $G_{liq}(x,y) = \r(x) \d(x,y) + \r(x) \r(y) h(x,y)$ and the two-point vertex functional of liquid theory 
$\G^{(2)}_{liq}(x,y) = \frac 1{\r(x)} \d(x,y) - c(x,y)$. Differentiating this equation with respect to the density we obtain the third order
Ornstein-Zernicke equation:
\beq \begin{split}
& \G^{(3)}_{liq}(x,y,z) = \\
& - \int_{x',y',z'} \!\!\!\!\!\!\!\! \G^{(2)}_{liq}(x,x') \G^{(2)}_{liq}(y,y') \G^{(2)}_{liq}(z,z') W^{(3)}_{liq}(x',y',z') \ ,
 \label{linkGamma3W3}
\end{split} \eeq
where $\G^{(3)}_{liq}$ is related to the third order direct correlation function via Eq.~(\ref{def_c}):
\beq
\G^{(3)}_{liq}(x,y,z) = - \frac 1{\r(x)^2} \d(x,y) \d(x,z) - c^{(3)}_{liq}(x,y,z) \ . 
\label{link_Gamma3_c3_liq}
\eeq
We will make repeated use of relations (\ref{OZ2_fieldtheory}--\ref{link_Gamma3_c3_liq}) in the next section.

\subsection{Replicated liquid theory}

We can now formulate in the same way the partition function of our replicated system, defining:
\beq
Z_m = \Tr e^{- \frac{\b}2 \underset{i,j,a,b}{\sum'} v_{ab}(x_i^a - x_j^b) + \underset{i,a}{\sum} \n_a(x_i^a)} \ ,
\eeq
where $v_{ab}$ is equal to $v$ for $a=b$, and $v_{ab}$ is a small attractive coupling when $a \ne b$. 
The prime on the summation sign $\sum'$ means that when $a=b$, the summation must exclude the case $i=j$, 
and the trace operation is now defined 
\mbox{as: $\Tr \bullet = \sum_{N=0}^\io \frac 1{N!^m} \int \prod_{a=1}^m \prod_{i=1}^N dx_i^a ~ \bullet$.} For the sake of 
simplicity, $\bfn$ in the following will denote the set of $m$ chemical potentials $\{\n_a\}_{a=1 \dots m}$, and $\bf{v}$ will 
denote the set of $m^2$ pair potentials $\{ v_{ab} \}_{a,b=1 \ldots m}$. Equivalently the family of $m$ average densities $\{ \r_a \}_{a=1 \ldots m}$ defined in 
Eq.~(\ref{def_roa}) will be denoted by $\bfr$ and the family of $m^2$ correlation functions $\{ h_{ab} \}_{a,b=1 \ldots m}$ will be denoted by 
$\bf{h}$. 

As for the non replicated liquid case, one can define the Legendre transform of the free-energy with respect to $\bfn$, to obtain an 
$m$-dependent Legendre transform $\G_{1,m}$:
\beq
\left\{ \begin{array}{ll}
& \displaystyle \G_{1,m}[\bfr] = \sum_a \int_x \r_a(x) \n_a^*[\bfr](x) - \ln Z_m[ \bfn^*] \ , \\
& \\
& \displaystyle  \left. \frac{\d \ln Z_m[\bfn]}{\d \n_a(x)} \right|_{\bfn^*} = \r_a(x) \ .
\end{array} \right.
\eeq

However, our goal here is to focus on the static order parameter $h_{ab}$ defined earlier. It is thus preferable to obtain a theory that explicitly 
depends on it in order to control approximations in terms of this quantity.
The key point is to notice that the pair potentials $v_{ab}$ are coupled to the two-point densities of the liquid $\r^{(2)}_{ab}$, defined by:
\beq\begin{split}
& \hro2_{ab}(x,y) = \hr_a(x) \hr_b(y) - \hr_a(x) \d_{ab} \d(x,y) \ , \\
& \r^{(2)}_{ab}(x,y) = \la \hro2_{ab}(x,y) \ra \ .
\end{split}\eeq
This quantity is trivially related to $h_{ab}$ by 
\beq
\r^{(2)}_{ab}(x,y) = \r_a(x) \r_b(y) \left[1 + h_{ab}(x,y) \right] \ ,
\label{link_h_ro2} 
\eeq
and we define it for convenience.
Indeed we remark that:
\beq\begin{split}
& Z_m = \Tr  e^{\Tr \frac{\hr^{(2)}_{ab}(x,y)}{2} (-\b v_{ab}(x,y)) + \Tr \hr_a(x) \n_a(x)} \ , \\
\Rightarrow & \frac{\delta \ln Z_m}{\delta (-\b v_{ab}(x,y))} = \frac 12 \r^{(2)}_{ab}(x,y) \ , \\
\Rightarrow & \frac{\delta \G_{1,m}[\bfr,\bf{v}]}{\delta \b v_{ab}(x,y)} = \frac 12 \r^{(2)}_{ab}(x,y) \ .
\end{split}\eeq
This relation explains why it will be easier to manipulate the properties of the Legendre transform that we define below in terms of $\r^{(2)}$ than in terms of $h$, avoiding 
unnecessary density factors in the calculations.
Also for later convenience, we define the propagator and two-point vertex function:
\beq \begin{split}
& G_{ab}(x,y) = \r_a(x) \d_{ab} \d(x,y) + \r_a(x) \r_b(y) h_{ab}(x,y) \ , \\
& \G^{(2)}_{ab}(x,y) = \frac1{\r_a(x)} \d_{ab} \d(x,y) - c_{ab}(x,y) \ ,
\end{split} \eeq
where $c_{ab}$ is the generalization to mixtures of the direct correlation function. Again we have 
the OZ relation, i.e. that $\Tr \G^{(2)}_{ac}(x,z) G_{cb}(z,y) = \d_{ab} \d(x,y)$, or equivalently
$h_{ab}(x,y) = c_{ab}(x,y) - \sum_c \int_z h_{ac}(x,z) \r_c(z) c_{cb}(z,y)$.

Defining $\G_m$ the Legendre transform of $\G_{1,m}$ with respect to $\b \bf{v}$ we get:
\beq
\left\{ \begin{array}{ll}
& \displaystyle \G_m[\bfr,{\bf h}] = - \frac 1 2 \sum_{a,b}  \int_{x,y} \r^{(2)}_{ab}(x,y) \b v_{ab}^*(x,y) \\
& \phantom{\G_m[\bfr,{\bf h}] =} + \G_{1,m}[\bfr,{\bf v}^*] \ , \\
& \\
& \displaystyle \left. \frac{\delta \G_{1,m}[\bfr,{\bf v}]}{\delta \b v_{ab}(x,y)} \right|_{{\bf v}^*} = \frac 1 2 \r^{(2)}_{ab}(x,y) \ ,
\end{array} \right.
\eeq
where we have written $\G_m$ as a function of $h$ since $\r^{(2)}$ and $h$ are simply related.

The functional $\G_m$, when evaluated at the true correlation function, coincides with $\G_{1,m}$, but now a stationary 
principle allows to derive self-consistent equations for the correlation function, because as a consequence of Legendre
transforms properties we have:
\beq
\begin{split}
\frac{\delta \G_m[\bfr,{\bf h}]}{\delta h_{ab}(x,y)} & = \r_a(x) \r_b(y) \frac{\delta \G_m[\bfr,{\bf h}]}{\delta \r^{(2)}_{ab}(x,y)} \\
& = - \frac 12 \r_a(x) \r_b(y) \b v_{ab}^*[{\bf h}](x,y) \ .
\end{split}
\eeq
We know that the values $v_{ab}^*$ of the pair potentials that lead to the real correlation function are $v_{ab}$,
so that the real correlation functions $h_{ab}$ are solutions to the 
self-consistent equation:
\beq
\frac{\delta \G_m[\bfr,{\bf h}]}{\delta h_{ab}(x,y)} = - \frac 12 \r_a(x) \r_b(y) \b v_{ab}(x,y) \ . 
\label{stat_principle_h}
\eeq
Now performing an approximation for $\G_m$ and using Eq.~(\ref{stat_principle_h}) leads to an approximate value of 
$h_{ab}$, and evaluating the approximated $\G_m$ at this value of $h_{ab}$ lead to an approximate free-energy that is 
consistent with the approximate $h_{ab}$ obtained. Finally, as described in Section~\ref{sec2}, we set the inter-replica
potential to zero, 
which distinguishes between the case $a=b$, where $v_{ab} = v$, and the case $a \ne b$, where 
$v_{ab}$ is now set to zero. Note that we set the inter-replica potential to zero only after taking the 
Legendre transform. This ensures that the thermodynamic limit has been taken, and will give rise to non trivial inter-replica correlations. 
We look for persisting inter-replica correlations, which will indicate the glass phase~\cite{MP96}.

\subsection{Morita \& Hiroike functional}

Morita \& Hiroike~\cite{MH61} showed that the functional $\G_m$ can be written as:
\beq
\G_m[\bfr,{\bf h}] = \G_{Id}[\bfr,{\bf h}] + \G_{Ring}[\bfr,{\bf h}] + \G^{2PI}[\bfr,{\bf h}] \label{morita} \ ,
\eeq
\begin{widetext}
where:
\beq \begin{split}
\G_{Id}[\bfr,{\bf h}] & = \sum_a \int_x \r_a(x) \left[ \ln \r_a(x) - 1 \right] \\
& ~ + \frac 12 \sum_{a,b} \int_{x,y} \r_a(x) \r_b(y) \left[ (1+h_{ab}(x,y)) \ln(1+h_{ab}(x,y)) - h_{ab}(x,y) \right] \ , \\
\G_{Ring}[\bfr,{\bf h}] & = \frac 12 \sum_{n \ge 3} \frac{(-1)^n}{n} \Tr \r_{a_1}(x_1) h_{a_1 a_2}(x_1,x_2) \cdots \r_{a_n}(x_n) h_{a_n a_1}(x_n,x_1) \ ,
\label{gamma_ring}
\end{split} \eeq
\end{widetext}
and $\G^{2PI}[\bfr,{\bf h}]$ is the sum of all 2PI diagrams, which are usual Mayer diagrams composed of black nodes $\r_a(x)$ and 
links $h_{ab}(x,y)$, such that when two 
links are removed from the diagrams, it does not disconnect in two separate parts. $\G_{Ring}$ is the sum of all ring diagrams. The 
Hyper-Netted-Chain (HNC)
approximation~\cite{MH59}, and other performant approximations of liquid theory start from this exact functional. This result is 
exactly what
we wanted: we have now an explicit functional of our static order parameter. Note that we have expressed everything as functionals of $h$ instead of $\r^{(2)}$ because it simplifies many 
calculations, but the natural independant variables are $\r$ and $\r^{(2)}$. Since $h$ depends of $\r$ and $\r^{(2)}$ through Eq.(\ref{link_h_ro2}), care must be taken when performing functional differentiations with 
respect to $\r$: they are meant at $\r^{(2)}$ fixed and not $h$ fixed.

Before turning to the expansion of this quantity in 
powers of the order parameter, we first review 
the results obtained within the HNC approximation, starting from the above Gibbs free-energy, since it is at the core of our 
expansion.

\subsection{HNC approximation for the replicated free-energy}

HNC amounts to discard all the 2PI diagrams, giving an analytic expression for the free-energy as 
a functional of ${\bf h}$ which is just the sum of the first two terms in Eq.~(\ref{gamma_ring}):
\beq
\G_m \approx \G_m^{HNC} = \G_{Id} + \G_{Ring} \ .
\eeq 
With this approximate functional $\G^{HNC}_m$, we obtain self consistent 
equations for the $h_{ab}$ by making use of Eq.~(\ref{stat_principle_h}). The sum of ring diagrams, when 
differentiated with respect to $h_{ab}$, i.e. when cutting a link, gives a sum of open chains, which is equal to 
$\bf{h}-\bf{c}$. 
Indeed, rewriting the generalization of the Ornstein-Zernike equation~(\ref{OZ_2}) to multicomponent mixtures in direct
space we have:
\beq
c_{ab}(x,y) = h_{ab}(x,y) - \sum_c \int_{z} h_{ac}(x,z) \r_c(z) c_{cb}(z,y) \ , \label{OZ2rep}
\eeq
which can be solved iteratively with respect to $h$, to give for all values of $a$ and $b$:
\begin{align}
& c_{ab}(x,y) = h_{ab}(x,y) \label{OZ2rep_expanded} \\
& + \sum_{n=1}^\io (-1)^{n} h_{a a_1}(x,x_1) \r_{a_1}(x_1) \cdots \r_{a_n}(x_n) h_{a_n b}(x_n,y) \nonumber \ ,
\end{align}
where summation over repeated indices and integration over repeated positions has been assumed.
Comparison with Eq.~(\ref{gamma_ring}) shows that $h_{ab}-c_{ab}$ is indeed exactly the derivative of $\G_{Ring}$ with respect to $h_{ab}$:
\beq
\frac{\d \G_{Ring}[\bfr,{\bf h}]}{\d h_{ab}(x,y)} = -\frac 12 \r_a(x) \r_b(y) \left[ h_{ab}(x,y) - c_{ab}(x,y) \right] \ .
\label{dGring_dh}
\eeq
Now using the variational principle stated in Eq.~(\ref{stat_principle_h}) along with the prescription described in Section~\ref{sec2}, 
we get a set of $m^2$ self consistent equations, which are different depending on whether we consider them for equal or different 
replica indices, due to the explicit breaking of the replica symmetry with the pair potentials $v_{ab}$:
\beq \begin{split}
\ln (1 + h_{aa}(x,y)) & = - \b v(x,y) + h_{aa}(x,y) - c_{aa}(x,y) \ , \\
& \\
\ln (1 + h_{ab}(x,y)) & = h_{ab}(x,y) - c_{ab}(x,y) ~ \text{ for } a \ne b \label{RHNC} \ .
\end{split} \eeq
These equations must be supplemented with the replicated Orstein-Zernike equations (\ref{OZ2rep}) that read in Fourier
space:
\beq \begin{split}
c_{aa}(k) = & h_{aa}(k) - \r h_{aa}(k) c_{aa}(k) - \r \sum_{c \ne a} h_{ac}(k) c_{ca}(k) \ , \\
c_{ab}(k) = & h_{ab}(k) - \r h_{aa}(k) c_{ab}(k) - \r h_{ab}(k) c_{bb}(k)  \\ 
& - \r \sum_{c \ne a \ne b} h_{ac}(k) c_{cb}(k) \ . \label{OZ2rep_fourier}
\end{split} \eeq 
In order to follow the prescription of Section~\ref{sec2}, we must make the free energy analytic in $m$, then take the limit 
$m \to 1$. In order to do that, we will assume replica symmetry (RS), i.e. that:
\beq
\left\{\begin{array}{ll}
\displaystyle h_{aa}(x,y) = h(x,y) & \quad \forall a \ , \\
\displaystyle h_{ab}(x,y) = \th(x,y) & \quad \forall a \ne b \ .
\end{array} \right. \label{1RSB}
\eeq
This ansatz allows to perform the summations over replica indices in Eqs.(\ref{RHNC}--\ref{OZ2rep_fourier}) 
\beq \begin{split}
\left\{ \begin{array}{ll}
c(k) = & h(k) - \r h(k) c(k) - \r (m-1) \th(k) \tc(k) \ , \\
& \\
\tc(k) = & \th(k) - \r h(k) \tc(k) - \r \th(k) c(k)  \ . \\
& - \r (m-2) \th(k) \tc(k)
\end{array} \right. 
\label{OZ2_interm}
\end{split} \eeq
We can now perform the $m \to 1$ limit, to obtain two sets of equations.
In the first set, we see that the functions $h$ and $c$ decouple from $\th$ and $\tc$ to give the usual liquid theory Ornstein-Zernike 
equation. Combined with the first equation of (\ref{RHNC}), we get the HNC approximation of liquid theory:
\beq
\left\{ \begin{array}{ll}
& \ln (1 + h(x,y)) = - \b v(x,y) + h(x,y) - c(x,y) \ , \\
& \\
& c(k) = h(k) - \r h(k) c(k) \ .
\end{array} \right. \label{HNC_diag}
\eeq
The second set OZ equations in (\ref{OZ2_interm}) specify the functions $\th$ and $\tc$, given $h$ and $c$. Combined with the self-consistent equation on the off-diagonal pair correlation $\th$, we get the replicated-HNC (RHNC) equations:
\beq
\left\{ \begin{array}{ll}
& \ln (1 + \th(x,y)) = \th(x,y) - \tc(x,y) \ , \\
& \\
& \tc(k) = \th(k) - \r h(k) \tc(k) - \r \th(k) c(k) + \r \th(k) \tc(k) \ .
\end{array} \right. \label{HNCrep_offdiag}
\eeq
We can rewrite the second of these equations as:
\beq
\tc(k) = \frac{(1 - \r c(k)) \th(k)}{1+\r h(k) - \r \th(k)} \ ,
\label{OZ2rep_offdiag_m=1}
\eeq
and use the definition of the structure factor Eq.~(\ref{def_Sk}), the Ornstein-Zernike equation Eq.~(\ref{OZ_2}) and the link
between $h_{ab}$ for $a \ne b$ with the non-ergodicity factor Eq.~(\ref{link_f_hab}) to obtain:
\beq
\tc(k) = \frac 1{\r S(k)} \frac{f(k)}{1-f(k)} \ . 
\eeq
Since $\tc$ and $\th$ are both related to $f$, we see that Eq.~(\ref{HNCrep_offdiag}) is a self-consistent equation on the 
non-ergodicity parameter, which reads
\beq
\frac 1{\r S(k)} \frac{f(k)}{1-f(k)} = \FF\big[ \th-  \ln (1 + \th)  \big](k) \ ,
\eeq
where $\FF$ denotes a Fourier transform and $\th$ is expressed in terms of $f$ by Eq.~(\ref{link_f_hab}).
If we were to make an expansion of the r.h.s. of this equation in powers of $f(k)$ (hence of $\th$), 
we would get, at lowest order:
\beq
\frac{f(k)}{1-f(k)} = \frac {S(k)}{2 \r} \int_q S(q) S(k-q) f(q) f(k-q) \ . \label{eq_f_HNC}
\eeq
This form of self consistent equation is very reminiscent of the Mode-Coupling result that read~\cite{bookGotze}:
\begin{widetext}
\beq\begin{split}
& \frac{f(k)}{1-f(k)} \!\! = \!\! \frac {S(k)}{2 \r} \!\! \int_q \!\! \MM_{\rm{MC}}(k,q) S(q) S(k-q) f(q) f(k-q) \ , \\
& \MM_{\rm{MC}}(k,q) = \frac{\r^2}{k^4} \left( k \cdot q c(q) + k \cdot(k-q) c(k-q) + \r k^2 c^{(3)}_{liq}(k,-q) \right)^2 \\
& \phantom{\MM_{\rm{MC}}(k,q)} = \left( \frac{k \cdot q}{k^2} \left[ 1 - \r c(q) \right] + \frac{k \cdot (k-q)}{k^2} \left[ 1 - \r c(k-q) \right] - \left[ 1 + \r^2 c^{(3)}_{liq}(k,-q) \right]  \right)^2 \\
& \phantom{\MM_{\rm{MC}}(k,q)} = \left( \frac{k \cdot q}{k^2} \r \G^{(2)}_{liq}(q) + \frac{k \cdot (k-q)}{k^2} \r \G^{(2)}_{liq}(k-q) + \r^2 \G^{(3)}_{liq}(k,-q) \right)^2 \ .
\label{MCT_kernel}
\end{split}\eeq
\end{widetext}
Note here the presence of the three-body direct correlation function of the liquid, even though it is usually neglected, since it has been shown 
to be negligible with respect to the other, two-body term~\cite{BGL89}, except for special cases~\cite{AGA11}.
However the RHNC result if expanded at this order in $h_{ab}$ is trivial (in the sense that the kernel $\MM=1$), and reminiscent of the result obtained by~\cite{ABL06}
by a dynamical field-theory calculation. Whether this is a coincidence or not is an open question at this stage. 
Nevertheless, we performed this expansion of the logarithmic term in Eq.~(\ref{HNCrep_offdiag}) only to show the striking similarities that
exist between replica calculations and usual Mode-Coupling ones.

\begin{figure}[t]
\includegraphics[width=8cm]{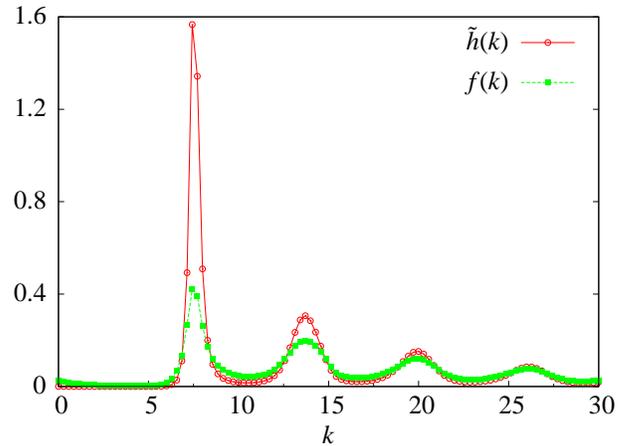}
\caption{Filled squares: Non-ergodicity factor of hard spheres at packing fraction $0.599670$ (i.e. at the transition), obtain from the replicated Hyper-Netted-Chain 
approximation Eq.~(\ref{HNCrep_offdiag}), combined with the liquid theory HNC approximation Eq.~(\ref{HNC_diag}). Open circles: the corresponding order parameter 
$\tilde{h}(k)$}
\label{fig:fk_RHNC}
\end{figure}

If we do not perform this truncation, and solve the RHNC approximation Eq.~(\ref{HNCrep_offdiag}) combined with the HNC approximation for the
liquid part Eq.~(\ref{HNC_diag}), we already find, {\it without assuming it}, the existence of a dynamical transition~\cite{MP96}: for hard spheres, 
for densities lower than $\varphi_d \approx 0.599$ (where the packing fraction $\varphi = \r \pi/6$), 
the solution to Eq.~(\ref{HNCrep_offdiag}) is always $h_{ab} = 0$, i.e. a liquid phase, whereas 
$h_{ab}$ discontinuously jumps to a non zero value for $\varphi \ge \varphi_d$, indicating the glass transition. 
This results is quite good, because the dynamical transition for three-dimensional hard spheres is estimated to
be around $\varphi_d \approx 0.57$. Note that Mode-Coupling Theory instead strongly underestimates the transition~\cite{Go99}.
We show in Fig. \ref{fig:fk_RHNC} the resulting 
non-ergodicity factor~\cite{PZ10} for a packing fraction $\varphi = 0.6$,
obtained by solving the set of equations (\ref{HNCrep_offdiag}).
This is the crucial problem of RHNC: $f(k)$ is quite far from the numerical results (see~\cite{MUP91,Go99}) which are instead well captured by 
the Mode-Coupling theory.

The goal of this paper is to demonstrate that the order parameter $h_{ab}$ with $a \ne b$ can be used as an organizing device for the theory in order 
to gradually incorporate higher-order correlations of the liquid into the replica result. Of course, we see in Fig. \ref{fig:fk_RHNC} that the order 
parameter is not a small quantity, and thus an expansion in powers of $h_{ab}$ is not {\it a priori} justified. Note however that RHNC already re-sums
an infinite number of diagrams containing arbitrary numbers of $h_{ab}$ links, which maybe explains its ability to predict a transition towards a 
non-small value of the order parameter. Our purpose here is to build from RHNC and incorporate more diagrams, and we will see that even keeping the lowest order 
correction already provides some improvement over RHNC.

\subsection{Improvements over the liquid quantities}

\begin{figure}[t]
\includegraphics[width=8cm]{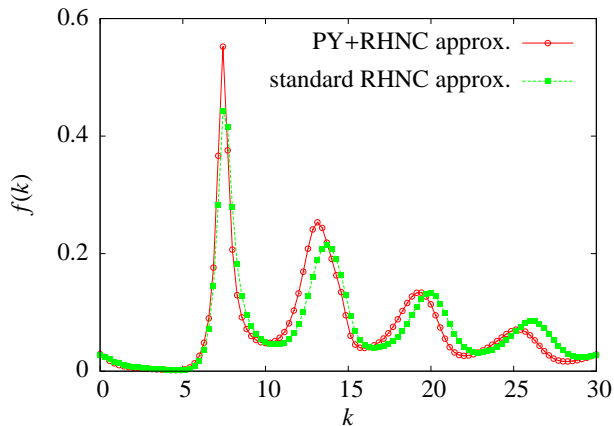}
\caption{Non-ergodicity factor as a function of the wave vector at packing fraction $0.6$, for the standard replicated HNC calculation, and from the 
combination of PY approximation for the diagonal part and RHNC for the off-diagonal correlation.}
\label{fig:compRHNC_PYRHNC_fk}
\end{figure}

Before turning to the study of the expansion in powers of the order parameter, a remark is in order. We have seen that in the $m=1$ limit, the liquid quantities decouple
from the inter-replica correlations: Eq.~(\ref{HNC_diag}) is in fact independent from the off-diagonal correlations. 
This allows us to use any liquid theory approximation to evaluate liquid quantities that appear in our equations. 
Formally, this is justified by writing the Gibbs free-energy $\Gamma_m$ in Eq.~(\ref{morita}) as:
\beq
\G_m[\bfr,{\bf h}] = \G_{m}^{HNC}[\bfr,{\bf h}] + \G^{2PI}_{liq}[\bfr,{\bf h}] + \G^{2PI}_{glass}[\bfr,{\bf h}] \ , \label{glass_liquid_2PI}
\eeq
where $\G^{2PI}_{liq}$ is the sum of all 2PI diagrams that do not contain any $h_{ab}, a \ne b$ links. We can now use the variational principles, which will give the full
liquid correlation function for the $a = b$ components. For a given approximation of $\G^{2PI}_{glass}$, we will obtain a self consistent equation for the $a \ne b$ components.
Neglecting altogether $\G^{2PI}_{glass}$, we recover the set of equations Eqs.(\ref{HNCrep_offdiag}), in which $c$ and $h$ are now the full liquid correlation functions. 
Of course, inside the glass phase, the liquid quantities cannot be obtained numerically or experimentally, so that such a full re-summation is useless. Instead, we need to have
an approximation that can be extrapolated from the liquid phase.
For example we can choose the PY approximation for the liquid quantities.
Choosing to work at a packing fraction (defined as $\varphi = \p \r / 6$) $\varphi = 0.6$, which is expected to be above the dynamical transition,
we solved the RHNC equation for the off-diagonal part, Eq.~(\ref{HNCrep_offdiag}) 
with the PY direct correlation function as an 
input for the diagonal part, and found that this brings about a little improvement over standard RHNC results. 
It is known that the PY approximation gives a less important underestimation of the peak of $S(k)$, 
which is the main ingredient that leads to the glass transition.
In standard RHNC as in the PY version of it, we find a transition from a liquid state at $\varphi < \varphi_d$ to a glass state at $\varphi \ge \varphi_d$, where 
the self-consistent equation on ${\tilde h}$ admits a non-zero solution. The value of the critical density is shifted downwards from 
$\varphi_c \approx 0.599$ to $0.591$ by the use of PY approximation, which is an improvement, even if modest.
We show the resulting non-ergodicity factor in the two sets of approximations in Fig. \ref{fig:compRHNC_PYRHNC_fk}. 
Even though use of the PY approximation gives a slightly larger non ergodicity factor, the result is still very small when compared to simulation and 
experimental data, where $f$ is much closer to $1$ at small wave vectors.
We conclude that the source of the problem in the static approach is not the diagonal part of the liquid. We should therefore seek for a way of improving
the equation for the off-diagonal correlation, Eq.~(\ref{HNCrep_offdiag}).

\subsection{Systematic expansion in powers of $\tilde{h}$}

To go further than the replicated Hyper-Netted-Chain approximation, we perform an expansion of $\G^{2PI}_{glass}$ in powers of $\tilde{h}$:
\begin{widetext}
\beq
\G^{2PI}_{glass}[\bfr,{\bf h}] = \sum_{n=1}^\io \Tr' \frac 1{n!} \left. \frac{\d^n \G^{2PI}_{glass}[\bfr,{\bf h}]}{\d h_{a_1 b_1}(x_1,y_1) \cdots \d h_{a_n b_n}(x_n,y_n)} \right|_{\th=0} h_{a_1 b_1}(x_1,y_1) \cdots h_{a_n b_n}(x_n,y_n) \ , \label{taylor_2PIglass}
\eeq
\end{widetext}
where $a_i \ne b_i$ $\forall i$, and we have underlined the fact that the zeroth order term is absent since the full functional must vanish in the liquid phase, where 
$\G^{2PI}_{glass} = 0$ by construction.

\begin{figure}[ht]
\begin{center}
\fcolorbox{white}{white}{
 \begin{picture}(174,129) (1,1)
   \SetWidth{1.0}
   \SetColor{Black}
   \COval(22,66)(63,20)(0){Black}{White}
   \COval(152,65)(63,20)(0){Black}{White}
   \Photon(38,104)(133,105){4}{5}
   \Photon(41,66)(131,66){4}{5}
   \Photon(37,29)(135,28){4}{5}
   \Vertex(136,26){5.385}
   \Vertex(132,64){5.385}
   \Vertex(135,103){5.385}
   \Vertex(37,31){5.385}
   \Vertex(41,69){5.385}
   \Vertex(36,105){5.385}
   \Text(17,64)[lb]{\Large{\Black{$a$}}}
   \Text(150,62)[lb]{\Large{\Black{$b$}}}
\end{picture}
}
\end{center}
\caption{Diagrams that contribute to the free-energy at order $\tilde{h}^3$. A wiggly line joining two replica indices $a$ and $b$ 
is a $h_{ab}$, with $a \ne b$ function, a black dot attached to a zone with replica index $a$ is an integration point weighted by a density 
factor $\r_a$.}
\label{diagrams}
\end{figure}
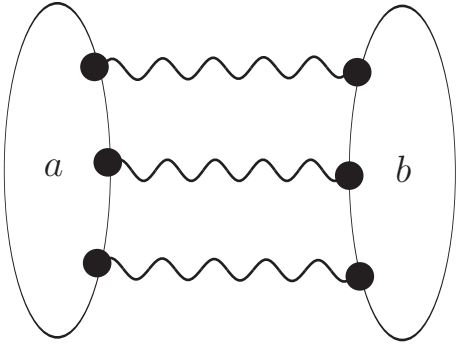

Now consider the ``glassy" 2PI diagrams in Eq.~(\ref{glass_liquid_2PI}). We can show that these diagrams must contain at least three
${\tilde h}$ lines: this idea was already used in the ``weak glass'' expansion of~\cite[Appendix A3]{BB09}. 
Indeed, a $\th$ link joins two nodes that have different replica indices, say $a$ and $b$. All the nodes connected to the 
$a$ node by a path of $h$ links must also have replica index $a$, and the same applies for the $b$ node. 
Thus all $\th$ links are nodal links: they separate the diagram in two parts, each of it has a different replica index. If a 
2PI diagram would contain one or two $\th$ links, then differentiating once or twice with respect to $\th$ would cut the 
diagram in two, which is in contradiction with the fact that the diagram is 2PI. Thus we proved that all 2PI diagrams contain 
either zero or three or more $\th$ links.
Moreover, a diagram that contains three $\th$ links can have at most six parts composed of $h$ links and $\r$ nodes
that all have the same replica index. A little reflexion shows that the only possibility to construct such diagram and 
make it 2PI is the one pictured in Fig.\ref{diagrams}.

From this analysis, we showed that the non HNC terms in the free-energy Eq.~(\ref{morita}) can be organized as a series in powers of $\tilde{h}$, 
the first term of which are $\OO(\tilde{h}^3)$. This last observation is crucial, since it will allow us to make the computation tractable.

In the following we will need to 
distinguish the derivatives with respect to the density and with respect to the correlation 
functions, as well as derivatives with respect to chemical potentials and with respect to pair potentials. We define:
\beq \begin{split}
\G^{(1,0)}_a(x_1) & = \frac{\delta \G_m[\bfr,{\bf h}]}{\delta \r_a(x_1)} \ , \\
\G^{(0,1)}_{ab}(x_1,y_1) & = \frac{\delta \G_m[\bfr,{\bf h}]}{\delta h_{ab}(x_1,y_1)} \ , \\
\G^{(1,1)}_{a,cd}(x_1;x_2,y_2) & = \frac{\delta^2 \G_m[\bfr,{\bf h}]}{\delta \r_a(x_1) \delta h_{cd}(x_2,y_2)} \ ,
\end{split} \eeq
and so on. 
We will also need to define derivatives of $\ln Z_m$ with respect to the chemical potentials and pair potentials:
\beq \begin{split}
W^{(1,0)}_a(x_1) & = \left. \frac{\delta \ln Z_m[\bfn,{\bf v}]}{\delta \n_a(x_1)} \right|_{\bfn^*,{\bf v}^*} \ ,  \\
W^{(0,1)}_{ab}(x_1,y_1) & = \left. \frac{\delta \ln Z_m[\bfn,{\bf v}]}{\delta (- \b v_{ab}(x_1,y_1))} \right|_{\bfn^*,{\bf v}^*} \ ,  \\
W^{(1,1)}_{a,cd}(x_1;x_2,y_2) & = \left. \frac{\delta^2 \ln Z_m[\bfn,{\bf v}]}{\delta \n_a(x_1) \delta (- \b v_{cd}(x_2,y_2))} 
\right|_{\bfn^*,{\bf v}^*} \ ,
\end{split} \eeq
and so on. Because of our analysis, we see that the $n=1$ and $n=2$ terms in  Eq.~(\ref{taylor_2PIglass}) are necessary 
zero. In the following, we will calculate the first non-zero term in this expansion, the third-order term. In practice, we found it
easier to calculate the third derivative of the total free energy, and substract from it the third order term of the RHNC free-energy, i.e.
we calculated, with the above notations:
\beq \begin{split}
\G^{(0,3)}_{ab,cd,ef} & (x_1,y_1;x_2,y_2;x_3,y_3) 
= \\ 
& \frac{\d^3 \left[ \G_m^{HNC}[\bfr,{\bf h}] + \G^{2PI}_{glass}[\bfr,{\bf h}] \right]}{\d h_{ab}(x_1,y_1) \d h_{cd}(x_2,y_2) \d h_{ef}(x_3,y_3)} \ ,
\end{split}
\eeq
with $a \ne b$, $c \ne d$ and $e \ne f$.
The third order term of the RHNC free-energy will be simply calculated from Eqs.(\ref{morita}--\ref{gamma_ring}).

\section{Treatment at third order in the order parameter}
\label{sec4}

In order to make progress, we must find a way to evaluate the third derivative of the 
free-energy with respect to the correlation function. We can make use of the properties 
of the Legendre transform to do so. The only difficulty is that we performed two 
Legendre transformations, with respect to different objects, which renders difficult the 
use of properties such as the one written in Eq.~(\ref{linkGamma3W3}). 

\subsection{Third-order Ornstein-Zernicke relation}

Considering, for simplicity, a discretized version of our theory, we have that $\bfr$ is a $m$ by $M$ matrix, where $M$ is 
the number of points of the underlying lattice, and ${\bf h}$ is a $m$ by $m$ by $M$ by $M$ object, and the same applies 
to $\bfn$ and ${\bf v}$. We can write the two pairs $\bfr,{\bfr}^{(2)}$ and $\bfn,{\bf v}$ in two (big) vectors:
\beq \begin{split}
& \Psi \equiv \left( \bfr , \frac 12 \bfr^{(2)} \right) \ , \\
& \Phi \equiv \left( \bfn , {\bf -\b v} \right) \ ,
\end{split} \eeq
with the convention that if $i > mM$ an index $i$ of the vector $\Psi$ or $\Phi$ must be understood as a group of two spatial coordinates $(x_i,y_i)$ and 
two replica indices $(a_i,b_i)$, but if $i \le mN$, it must be understood as one spatial coordinate and one replica index.
The double Legendre transform $\G_m$ can then be written as:
\beq
\left\{ \begin{array}{ll}
& \G_m[\Psi] = \Tr \Phi^* \Psi - \ln Z_m[\Phi^*] \ , \\
& \\
& \displaystyle \text{with } \Phi^* \text{ such that } ~ \left. \frac{\d \ln Z_m[\Phi]}{\d \Phi} \right|_{\Phi^*} = \Psi \ .
\end{array} \right.
\eeq
We have by definition of the Legendre transform:
\beq
\Phi^*_1 = \frac{\d \G_m[\Psi]}{\d \Psi_1} \ .
\eeq
We can perform a functional derivative of this equation with respect to $\Psi_2$ to get:
\beq \begin{split}
& \left( \left. \frac{\d^2 \ln Z_m[\Phi]}{\d \Phi_1 \d \Phi_2} \right|_{\Phi^*} \right)^{-1} = \frac{\d^2 \G_m[\Psi]}{\d \Psi_1 
\d \Psi_2} \ , \\
\Leftrightarrow ~ & \d_{12} = \Tr \left( \frac{\d^2 \G_m[\Psi]}{\d \Psi_1 \d \Psi_4} \left. \frac{\d^2 \ln Z_m[\Phi]}{\d \Phi_4 
\d \Phi_2} 
\right|_{\Phi^*} \right) \ . \label{W2_G2_formal}
\end{split} \eeq
Computing a third derivative with respect to $\Psi_3$, we get formally:
\beq \begin{split}
& 0 = \Tr \left( \frac{\d^3 \G_m[\Psi]}{\d \Psi_1 \d \Psi_3 \d \Psi_4} \left. \frac{\d^2 \ln Z_m[\Phi]}{\d \Phi_4 \d \Phi_2} 
\right|_{\Phi^*} \right)  \\
& + \Tr \left( \frac{\d^2 \G_m[\Psi]}{\d \Psi_1 \d \Psi_4} \frac{\d^2 \G_m[\Psi]}{\d \Psi_3 \d \Psi_5} 
\left. \frac{\d^3 \ln Z_m[\Phi]}{\d \Phi_5 \d \Phi_4 \d \Phi_2} \right|_{\Phi^*} \right) \ , \\
& 
\end{split} \eeq
And multiplying through by a second derivative of $\G_m$ and using Eq.~(\ref{W2_G2_formal}), we get:
\begin{align}
& \frac{\d^3 \G_m[\Psi]}{\d \Psi_1 \d \Psi_2 \d \Psi_3} = \label{W3_G3_formal} \\ 
& - \Tr \left( \frac{\d^2 \G_m[\Psi]}{\d \Psi_1 \d \Psi_{1'}} 
\frac{\d^2 \G_m[\Psi]}{\d \Psi_2 \d \Psi_{2'}} \frac{\d^2 \G_m[\Psi]}{\d \Psi_3 \d \Psi_{3'}} 
\left. \frac{\d^3 \ln Z_m[\Phi]}{\d \Phi_{1'} \d \Phi_{2'} \d \Phi_{3'}} \right|_{\Phi^*} \right) \ , \nonumber
\end{align}
This is a standard equation that can be found in any textbook on field theory, for example in~\cite{zinn}, that expresses
the relation between vertex functions and propagators of the theory. It coincides with Eq.~(\ref{linkGamma3W3}) if the
vectors $\Psi$ and $\Phi$ contain only the first $mM$ components.

We are interested in the continuum space limit of this expression, evaluated in the liquid, and with $1,2$ and $3$ that 
are all greater than $mM$. With explicit spatial coordinates and replica indices and implicit summation and integration over repeated indices and variables, and after
replacing derivatives with respect to $\r^{(2)}$ with derivatives with respect to $h$, we get:
\begin{widetext}
\beq \begin{split}
\G^{(0,3)}_{ab,cd,ef}& (x_1,y_1;x_2,y_2;x_3,y_3)  \\
= & - \G^{(1,1)}_{ab,a'}(x_1,y_1;x_1') \G^{(1,1)}_{cd,c'}(x_2,y_2;x_2') \G^{(1,1)}_{ef,e'}(x_3,y_3;x_3')
W^{(3,0)}_{a',c',e'}(x_1';x_2';x_3')  \\
& - \frac {2}{\r^2} ~ \G^{(0,2)}_{ab,a'b'}(x_1,y_1;x_2',x_3') \G^{(1,1)}_{cd,c'}(x_2,y_2;x_2') \G^{(1,1)}_{ef,e'}(x_3,y_3;x_3')
W^{(2,1)}_{a'b',c',e'}(x_1',y_1';x_2';x_3')  \\
& - \text{ two permutations } \{(a,b);(x_1,y_1)\} \leftrightarrow \{(c,d);(x_2,y_2)\} \leftrightarrow \{(e,f);(x_3,y_3)\}  
\label{derivee3_gamma}  \\
& - \frac {4}{\r^{4}} ~ \G^{(1,1)}_{ab,a'}(x_1,y_1;x_1') \G^{(0,2)}_{cd,c'd'}(x_2,y_2;x_2',y_2') 
\G^{(0,2)}_{ef,e'f'}(x_3,y_3;x_3',y_3') W^{(1,2)}_{a',c'd',e'f'}(x_1';x_2',y_2';x_3',y_3')  \\			
& - \text{ two permutations } \{(a,b);(x_1,y_1)\} \leftrightarrow \{(c,d);(x_2,y_2)\} \leftrightarrow \{(e,f);(x_3,y_3)\} 
  \\
& - \frac {8}{\r^{6}} ~ \G^{(0,2)}_{ab,a'b'}(x_1,y_1;x_1',y_1') \G^{(0,2)}_{cd,c'd'}(x_2,y_2;x_2',y_2') 
\G^{(0,2)}_{ef,e'f'}(x_3,y_3;x_3',y_3') W^{(0,3)}_{a'b',c'd',e'f'}(x_1',y_1';x_2',y_2';x_3',y_3')  \ . 
\end{split}\eeq
\end{widetext}
We note that this expression is correct independantly of the value of $m$, the replica ansatz chosen, and the value of the fields 
$\bfr$ and $\bfh$ chosen. However three simplifications will occur: we are performing a Taylor expansion in powers of $h_{ab}$, 
with $a \ne b$, so that the indices $a,b,c,d,e$ and $f$ in Eq.~(\ref{derivee3_gamma}) must be chosen so that $a \ne b$, $c \ne d$ 
and $e \ne f$. Secondly we must evaluate the derivatives at zero off-diagonal correlation ($\th \to 0$ in the RS ansatz). Finally
we are interested only in the dynamical transition point, which is described by the $m \to 1$ limit. These three features will greatly 
simplify the calculation.

There are two types of objects that we need to compute in order to use this relation: cumulants of the microscopic densities
that are generated by the differentiation of $\ln Z_m$ with respect to $\n_a$ and $v_{ab}$, and second derivatives of
$\G_m$ with respect to $\r_a$ and $h_{ab}$. In the end, we want to evaluate these objects in the liquid phase where $\th$
is equal to zero. But we know that the free energy can be written as the HNC free energy plus
2PI contributions, that are $\OO(\th^3)$, i.e. contain more than three $\th$ links. Thus when taking one or 
two derivatives of the 2PI diagrams with respect to $\th$, they still will contain at least one $\th$ link, and will all cancel out 
when evaluated at zero $\th$. This proves that Eq.~(\ref{derivee3_gamma}) when evaluated in the liquid phase can be
computed by replacing the $\G_m$ functionals in the r.h.s. by $\G_m^{\rm{HNC}}$.

Thus the needed derivatives of $\G_m$ can be computed starting from Eq.~(\ref{morita}) by dropping the 2PI diagrams.
The only difficult term is the sum of ring diagrams. Its first derivative is given in Eq.~(\ref{dGring_dh}). Thus we have that:
\beq \begin{split}
\frac{\d \G_m^{HNC}[\bfr,\bfh]}{\d h_{ab}(x_1,y_1)} = \frac 12 \r_a(x_1) & \r_b(y_1) \left( \frac{}{} \ln (1 + h_{ab}(x_1,y_1)) \right. \\ 
& \left. - h_{ab}(x_1,y_1) + c_{ab}(x_1,y_1) \frac{}{} \right) 
\label{dGammaHNC_dh} \ .
\end{split}\eeq
In order to perform a second derivative with respect to $h$, one must resort to the expression of $c$ as a function of $h$ 
and $\r$ in Eq.~(\ref{OZ2rep_expanded}). We easily find:
\beq \begin{split}
\frac{\d c_{ab}(x_1,y_1)}{\d h_{cd}(x_2,y_2)} & = \r_c(x_2) \r_d(y_2) \\
& \times \left( \frac1{\r_a(x_1)} \d_{ac} \d(x_1,x_2) - c_{ac}(x_1,x_2) \right)  \\
& \times \left( \frac1{\r_b(y_1)} \d_{bd} \d(y_1,y_2) - c_{bd}(y_1,y_2) \right) \ .
\end{split} \eeq
In principle, this expression should be symmetrized with respect to a change of indices, but here everything will be traced in the end 
of the calculation, thus we can keep working with non-symmetrized quantities.
We obtain as a consequence:
\beq\begin{split}
& \frac{\d^2 \G_m^{HNC}[\bfr,\bfh]}{\d h_{ab}(x_1,y_1) \d h_{cd}(x_2,y_2)} = \\
& \frac 12 \r_a(x_1) \r_b(y_1) \left( \frac{1}{1 + h_{ab}(x_1,y_1)} -1\right) \d_{ab,cd}(x_1,y_1;x_2,y_2) \\
&+ \frac 12 \r_a(x_1) \r_b(y_1) \r_c(x_2) \r_d(y_2) \G^{(2)}_{ac}(x_1,x_2)\G^{(2)}_{bd}(y_1,y_2) \ .
\end{split}\eeq
And finally taking the limit $\th \to 0$ with $a \ne b$ we get:
\beq \begin{split}
\G^{(0,2)}_{ab,cd} & (x_1,y_1;x_2,y_2) \left. \right|_{liq}  \\
& = \frac 12 \r^4  \d_{ac} \d_{bd} \G^{(2)}_{liq}(x_1,x_2) \G^{(2)}_{liq}(y_1,y_2) \\
& \equiv \frac 12 \r^4 \d_{ac}\d_{bd} \G_{HNC}^{(0,2)}(x_1,x_2;y_1,y_2) \ .
\label{Gamma_02}
\end{split} \eeq
The other needed derivative is $\G^{(1,1)}$, which is calculated from Eq.~(\ref{dGammaHNC_dh}) by noting that differentiating 
$c$ with respect to the density, at $h$ fixed, simply gives a product of $c$ functions:
\beq
\frac{\d c_{ab}(x_1,y_1)}{\d \r_c(x_2)} = - c_{ac}(x_1,x_2) c_{bc}(y_1,x_2) \ .
\eeq
Now recall that derivatives with respect to density must be done at $\r^{(2)}$ fixed instead, so we use the chain rule:
\beq \begin{split}
& \left. \frac{\d c_{ab}(x_1,y_1)}{\d \r_c(x_2)} \right|_{\r^{(2)} \text{ cte}} = \left. \frac{\d c_{ab}(x_1,y_1)}{\d \r_c(x_2)} \right|_{h \text{ cte}} \\
& + \sum_{e,f} \int_{u,v} \left. \frac{\d h_{ef}(u,v)}{\d \r_c(x_2)} \right|_{\r^{(2)} \text{ cte}} \left. \frac{\d c_{ab}(x_1,y_1)}{\d h_{ef}(u,v)} \right|_{\r \text{ cte}} \ ,
\end{split} \eeq
to obtain the final result:
\beq \begin{split}
& \left. \frac{\d^2 \G_m^{HNC}[\bfr,\bfh]}{\d h_{ab}(x_1,y_1) \d \r_c(x_2)} \right|_{liq} = \\
& -\frac 1{2 S(k=0)} \left[ \frac{}{} \d_{ac} \G^{(2)}_{liq}(y_1,x_2) + \d_{bc} \G^{(2)}_{liq}(x_1,x_2) \right] \ ,
\label{Gamma_11}
\end{split} \eeq
where $S(k)$ is the structure factor of the liquid.

If we drop the space indexes, we can now perform the trace over replica indexes in
Eq.~(\ref{derivee3_gamma}), since all derivatives are delta functions with respect to replica indices. We obtain:
\beq \begin{split}
 \G^{(0,3)}_{ab,cd,ef} = & - \r^6 ~ \G^{(0,2)}_{HNC} \otimes \G^{(0,2)}_{HNC} \otimes \G^{(0,2)}_{HNC} \otimes W^{(0,3)}_{ab,cd,ef}  \\
& + \frac{\r^4}{2 S(0)} \G^{(2)}_{liq} \otimes \G^{(0,2)}_{HNC} \otimes \G^{(0,2)}_{HNC} \otimes W^{(1,2)}_{ab,cd,e} \\
& + \left\{ 2 \text{ perms.} \right\} \\
& - \frac{\r^2}{2 S(0)} \G^{(2)}_{liq} \otimes \G^{(2)}_{liq} \otimes \G^{(0,2)}_{HNC} \otimes W^{(2,1)}_{ab,c,e} \\
& - \left\{ 2 \text{ perms.} \right\} \\
& + \frac 1{8 S(0)} \G^{(2)}_{liq} \otimes \G^{(2)}_{liq} \otimes \G^{(2)}_{liq} \otimes W^{(3,0)}_{a,c,e} \ .
\label{Gamma_03_traced_replicas}
\end{split} \eeq
where $\otimes$ means space convolution with respect to two spatial indexes.

Finally, we only have left the task to compute the derivatives of the logarithm of the partition function with respect to pair 
potentials or chemical potentials. These terms, equivalently to the one written 
in Eq.~(\ref{W3_liquid}) for the liquid, are cumulants of microscopic one- or two-point densities, that are easily computed within the RS framework that we use here,
as we explain in the following.

\subsection{Replica symmetric structure of the theory}

We denote by $\la \bullet \ra$ the equilibrium average for the replicated system.
Once again, in the limit $v_{ab}(r)\to 0$, all replicas fall in the same state 
but are otherwise uncorrelated inside the state. Finally, we want to evaluate
all our averages in the liquid phase.
This leads to the following rule to compute the average $\la \bullet \ra$: one should
\begin{itemize}
\item
factorize the averages $\la \bullet \ra$ when they involve different replicas, and
\item
remove the replica indexes.
\item
replace $\la \bullet \ra = \la \bullet \ra_{liq}$
\end{itemize}
For instance, for any spatial argument, and for $a \neq b$, we have that following the prescription above
\beq
\la \hat\r_a \hat\r_b \ra =  \la \hat\r_a \ra \la \hat\r_b \ra  =  \la \hr \ra_{liq} \la \hr \ra_{liq} = \r^2 \ .
\eeq
Similarly, assuming that different letters denote different values of the indexes:
\beq \begin{split}
\la \hat\r_a \hat\r_a\hat\r_b \ra & =  
\la \hat\r_a\hat\r_a\ra \la \hat\r_b \ra 
 = \la \hr \hr \ra_{liq} \la \hr \ra_{liq} = \r (G_{liq} + \r^2) \ .
\end{split} \eeq
\beq\begin{split}
\la \hr^{(2)}_{ab} \hr^{(2)}_{ac} \ra & = \la \hr_a \hr_b ~ \hr_a \hr_c \ra = \la \hr_a \hr_a \ra \la \hr_b \ra \la \hr_c \ra \\
& = \la \hr \ra_{liq} \la \hr \ra_{liq} \la \hr \hr \ra_{liq} \\
& = \r^2 \left( G_{liq} + \r^2 \right) \ . \\
& 
\end{split}\eeq
We will thus obtain quantities that do not depend on replica indices anymore, 
allowing to sum over these indices, and finally take the $m \to 1$ limit. The free-energy will 
have an overall factor $m(m-1)$, and thus we will consider the free energy divided by $m(m-1)$.
Indeed, recalling that we will calculate a free-energy correction of the form:
\begin{widetext}
\beq \begin{split}
\d \G_m[\bfr,{\bf h}] & \equiv \frac 1{3!} \sum_{a \ne b} \sum_{c \ne d} \sum_{e \ne f} \int_{x_1,y_1,\cdots,x_3,y_3} \G^{(0,3)}_{ab,cd,ef}(x_1,y_1;x_2,y_2;x_3,y_3) h_{ab}(x_1,y_1) 
h_{cd}(x_2,y_2) h_{ef}(x_3,y_3) \\
& = -\frac {\r^6}{3!} \int_{x_1, \ldots, y_3} \left( \sum_{a \ne b, c \ne d, e \ne f} \G^{(0,3)}_{ab,cd,ef}(x_1,y_1;x_2,y_2;x_3,y_3) \right) \th(x_1,y_1) \th(x_2,y_2) \th(x_3,y_3) \label{derivee3_gamma_traced}
\end{split} \eeq
\end{widetext}
Now it is in order to remember that we will want to evaluate everything at $m=1$ at the end of the calculation. Everything will be proportional to $m-1$, thus we will first remove this
factor before evaluating. Afterwards, all terms that contain an additional factor $m-1$ will disappear. Now when we look at Eq.(\ref{Gamma_03_traced_replicas}), we see that
there are terms in which replica indices do not appear explicitely, for example in the term containing $W^{(3,0)}_{a,c,e}$, $b,d$ and $f$ do not appear. They are only constrained
to be different from their conjugate indices $a,c$ and $e$ respectively, thus when summing over all values of these indices, we will obtain three factors $m-1$, and the term will cancel in
the $m \to 1$ limit. This observation allow us to discard all terms but those containing $W^{(0,3)}$ in the limit $m \to 1$.

\subsection{Final calculation}

Before turning to the explicit evaluation of $W^{(0,3)}$, it is useful to remark that, within 
the RS structure that we have, we can parametrize its replica dependance in a simple way.

Take a matrix that depends on two pairs of replica indexes $M_{ab,cd}$, with $a \ne b$ 
and $c \ne d$ (we do not explicit the space indexes). Examination of the different possibilities for $a,b,c$ and $d$ shows that 
we have only three genuinely different possibilities:
\beq
M_{ab,cd} = \left\{ \begin{array}{ll}
& M_{ab,ab} \ , \\
& M_{ab,ac} \ , \\
& M_{ab,cd} \ .
\end{array} \right.
\eeq
This is a consequence both of the RS ansatz and of the symmetry of the functions 
with respect to permutations of indexes (and their associated space indexes).
This can be summarized in:
\beq \begin{split}
M_{ab,cd} = & 
M_1  \frac{\d_{ac}\d_{bd} + \d_{ad}\d_{bc}}2 \label{4replica_matrix_compact} \\
& + M_2 \frac{\d_{ac} + \d_{ad} + \d_{bc} + \d_{bd}}4 + M_3 \ , 
\end{split} \eeq
where $M_1,M_2$ and $M_3$ are related to the above terms by:
\beq
\left\{ \begin{array}{ll}
& M_1 = 2 \left[ M_{ab,ab} - 2 M_{ab,ac} + M_{ab,cd} \right] \ , \\
& M_2 = 4 \left[ M_{ab,ac} - M_{ab,cd} \right] \ , \\
& M_3 = M_{ab,cd} \ .
\end{array}\right.
\eeq
The quantity that we are interested in is a matrix that depends on three pairs of indexes. In this case there are 8 topologically different possibilities~\cite{TDP02}:
\beq \begin{split}
W^{(0,3)}_{ab,cd,ef} = \left\{ \begin{array}{ll}
W_1 = W_{ab,bc,ca} \ , \\
W_2 = W_{ab,ab,ab} \ , \\
W_3 = W_{ab,ab,ac} \ , \\
W_4 = W_{ab,ab,cd} \ , \\
W_5 = W_{ab,ac,bd} \ , \\
W_6 = W_{ab,ac,ad} \ , \\
W_7 = W_{ab,ac,de} \ , \\
W_8 = W_{ab,cd,ef} \ .
\end{array} \right. 
\end{split} \eeq
\vspace{2cm}
A relation like Eq.~(\ref{4replica_matrix_compact}) is again possible, but cumbersome,
and we do not write it explicitly because we do not need it.
By using the prescription for calculating averages of one and two-point densities 
described above, we can easily compute the $W_i$. We obtain:
\begin{widetext}
\beq \begin{split}
& W_1(x_1,\cdots,y_3) = \frac 1 8 \left[ \begin{array}{ll}
& G_{liq}(x_1,x_2)G_{liq}(y_1,x_3)G_{liq}(y_2,y_3) \\
& + \r^2 \left(G_{liq}(x_1,x_2)G_{liq}(y_1,x_3) + G_{liq}(x_1,x_2)G_{liq}(y_2,y_3) + G_{liq}(y_1,x_3)G_{liq}(y_2,y_3) \right) \end{array} \right] \ ,
\label{W_1}
\end{split}\eeq
\beq\begin{split}
& W_2(x_1,\cdots,y_3) = \frac 1 8 \left[ \begin{array}{ll} 
& W^{(3)}_{liq}(x_1,x_2,x_3) W^{(3)}_{liq}(y_1,y_2,y_3) \\
& + \r W^{(3)}_{liq}(x_1,x_2,x_3) \left( \r^2 + G_{liq}(y_1,y_2) + G_{liq}(y_1,y_3) + G_{liq}(y_2,y_3) \right) \\
& + \r W^{(3)}_{liq}(y_1,y_2,y_3) \left( \r^2 + G_{liq}(x_1,x_2) + G_{liq}(x_1,x_3) + G_{liq}(x_2,x_3) \right) \\
& + \r^2 \left( G_{liq}(x_1,x_2)G_{liq}(y_1,y_3) + G_{liq}(x_1,x_2)G_{liq}(y_2,y_3) + G_{liq}(x_1,x_3)G_{liq}(y_1,y_2) \right) \\
& + \r^2 \left( G_{liq}(x_1,x_3)G_{liq}(y_2,y_3) + G_{liq}(x_2,x_3)G_{liq}(y_1,y_2) + G_{liq}(x_2,x_3)G_{liq}(y_1,y_3) \right) 
\end{array} \right] \ ,
\label{W_2}
\end{split} \eeq
\beq \begin{split}
& W_3(x_1,\cdots,y_3) = \frac 18 \left[ \r^3 W^{(3)}_{liq}(x_1,x_2,x_3) + \r G_{liq}(y_1,y_2) \left( W^{(3)}(x_1,x_2,x_3) + \r G_{liq}(x_1,x_3) + \r G_{liq}(x_2,x_3) \right) \right] \ , 
\label{W_3}
\end{split} \eeq
\end{widetext}
\beq
W_5(x_1,\cdots,y_3) = \frac 18 \r^3 W^{(3)}_{liq}(x_1,x_2,x_3) \ ,
\label{W_5}
\eeq
\beq
W_6(x_1,\cdots,y_3) = \frac 18 \r^2 G_{liq}(x_1,x_2) G_{liq}(y_1,x_3) \ , 
\label{W_6}
\eeq
and we find $W_4 = W_7 = W_8 = 0$.
The factors $1/8$ come from the fact that a derivative of $\ln Z_m$ with respect to 
$-\b v$ gives $\r^{(2)}/2$ and not $\r^{(2)}$. Of course, the choice of spatial indexes
is arbitrary, and one can make any permutation, as long as it respects the symmetry of the functions. We can now perform the trace 
over replica indexes in Eq.~(\ref{derivee3_gamma_traced}), which will give an expression analytic in $m$. We omit the space indices
in the following, but they are recovered by considering all permutations of the space indices written in Eqs.~(\ref{W_1}--\ref{W_6}).
The number of terms are obtained by considering the number of possible ways to choose a particular arrangement of replica 
indices among $m$ indices. For example to construct a term contributing to $W_2$, one must first pick a value of $a$ ($m$ 
possibilities), then a different value of $b$ ($m-1$ possibilities) which shows that the trace of $W_2$ let appear $m(m-1)$ identical
terms (with the exact same space indices structure). In addition to this multiplicity, because of the invariance by permutation of 
indices (together with their corresponding space indices), we have that:
\beq
W^{(0,3)}_{ab,ab,ab} = W^{(0,3)}_{ab,ba,ab} = W^{(0,3)}_{ab,ab,ba} = W^{(0,3)}_{ab,ba,ba} \ ,
\eeq
and this will give four terms that have the same replica structure, but permutations of space indices. The multiplicity of $W_2$ is
thus $4m(m-1)$. Performing the same counting on all cubic masses, we obtain:
\begin{align}
\frac 1{m(m-1)}\sum_{a \ne b, c \ne d, e \ne f} \!\!\! & W^{(0,3)}_{ab,cd,ef} =  \\
& 4 W_2 + (m-2) \left[ 8 W_1 + 24 W_3 \right] \nonumber \\
& + (m-2)(m-3) \left[ 8 W_5 + 24 W_6 \right] \nonumber \\
& + \OO(m-1) \ . \nonumber
\end{align}
After taking the $m \to 1$ limit, we obtain:
\beq\begin{split}
& \lim_{m \to 1} \frac 1{m(m-1)} \sum_{a \ne b}\sum_{c \ne d}\sum_{e \ne f} W^{(0,3)}_{ab,cd,ef}  \\
& = 4 W_2 - 8 W_1 - 24 W_3 + 16 W_5 + 48 W_6  \\
& = \frac 12 W^{(3)}_{liq}(x_1,x_2,x_3) W^{(3)}_{liq}(y_1,y_2,y_3) \\
& \quad - G_{liq}(x_1,x_2)G_{liq}(y_1,x_3)G_{liq}(y_2,y_3)  \ .
\end{split}\eeq
Finally we can perform the convolution with the derivatives of the HNC free energy in Eq.~(\ref{derivee3_gamma_traced}), and by making repeated use of the second- and third-order OZ equations we obtain:
\beq\begin{split}
& \lim_{m \to 1} ~ \frac {\d \G_m[\bfr,{\bf h}]}{m(m-1)}  \nonumber  \\
& = - \frac {\r^6}{6} \int_{x_1,\cdots,y_3} \hspace{-0.7cm} V(x_1,\cdots,y_3) \th(x_1,y_1) \th(x_2,y_2) \th(x_3,y_3) \ ,  
\end{split} \eeq
\beq \begin{split}
& V(x_1,\cdots,y_3) = \frac 12 \G^{(3)}_{liq}(x_1,x_2,x_3) \G^{(3)}_{liq}(y_1,y_2,y_3) \label{result_with_HNCterm} \\
& \hspace{2.5cm} -\G^{(2)}_{liq}(x_1,x_2) \G^{(2)}_{liq}(y_1,x_3) \G^{(2)}_{liq}(y_2,y_3) \ . 
\end{split}\eeq
This third order correction includes the RHNC term, which is recovered by setting $c^{(3)}_{liq} = 0$ \cite{Iy84}, but which we can 
also recover by directly differentiating Eq.~(\ref{Gamma_02}). In any case we find:
\beq \begin{split}
V(x_1,\cdots,y_3) = & V^{HNC}(x_1,\cdots,y_3) \\
& + V^{2PI}(x_1,\cdots,y_3) \ ,
\end{split} \eeq
where we defined:
\beq \begin{split}
V^{HNC}(x_1,\cdots,y_3) = & - \G^{(2)}_{liq}(x_1,x_2) \G^{(2)}_{liq}(y_1,x_3) \G^{(2)}_{liq}(y_2,y_3) \\
& \hspace{-1cm} +\frac 1{2 \r^4} \d(x_1,x_2) \d(x_1,x_3) \d(y_1,y_2) \d(y_1,y_3) \ , 
\end{split} \eeq
which is the contribution coming from $\G_m^{HNC}$, and
\beq\begin{split}
V^{2PI}(x_1,\cdots,y_3) = & \frac 12 c^{(3)}_{liq}(x_1,x_2,x_3) c^{(3)}_{liq}(y_1,y_2,y_3) \\
& \hspace{-1cm} + \frac 1{2\r^2} \d(x_1,x_2)\d(x_1,x_3) c^{(3)}_{liq}(y_1,y_2,y_3) \ ,
\end{split}\eeq
which is the sought for contribution coming from $\G_m^{2PI}$.
We obtain finally:
\begin{widetext}
\beq \begin{split}
\left. \frac {\G_m[\bfr,{\bf h}]}{m(m-1)} \right|_{m=1} = \left. \frac{\G_m^{HNC}[\bfr,{\bf h}]}{m(m-1)} \right|_{m=1} & ~ - \frac{\r^6}{12} \int_{x_1, \cdots , y_3} c^{(3)}(x_1,x_2,x_3) c^{(3)}(y_1,y_2,y_3) \tilde{h}(x_1,y_1) \tilde{h}(x_2,y_2) \tilde{h}(x_3,y_3)  \\
& ~ - \frac{\r^4}6 \int_{x_1,x_2,x_3,y} c^{(3)}(x_1,x_2,x_3) \tilde{h}(x_1,y) \tilde{h}(x_2,y) \tilde{h}(x_3,y) \ .
\end{split} \eeq
\end{widetext}
This is the desired result: the next order term in the order-parameter expansion beyond
the RHNC approximation. From this approximation of the free-energy, we can make use 
of the variational principle Eq.~(\ref{stat_principle_h}) to obtain a closed equation on $\tilde{h}$:
\beq \begin{split}
& \tilde{c}(r) = \tilde{h}(r) - \ln [ 1 + \tilde{h}(r) ]  \\
& + \frac{\r^4}{2} \int_{r_1 \cdots r_4} \hspace{-0.5cm} c^{(3)}_{liq}(r,r_1,r_3) c^{(3)}(0,r_2,r_4) 
\tilde{h}(r_1,r_2) \tilde{h}(r_3,r_4)  \\
& + \frac{\r^2}2 \int_{r_1,r_2} \hspace{-0.3cm}c^{(3)}(r,r_1,r_2) \tilde{h}(r_1) \tilde{h}(r_2) \label{final_rspace} \\
& + \frac{\r^2}2 \int_{r_1,r_2} \hspace{-0.3cm}c^{(3)}(0,r_1,r_2) \tilde{h}(r-r_1) \tilde{h}(r-r_2) \ .  
\end{split} \eeq
which provides the first correction to Eq.~(\ref{HNCrep_offdiag}).
Using translational invariance as well as the invariance under permutation of the three variables of $c^{(3)}$, 
we get:
\beq\begin{split}
& c^{(3)}(r_1,r_2,r_3) = c^{(3)}(r_1-r_2;r_1-r_3) \equiv c^{(3)}(r;s) \ , \\
& \text{with } r = r_1-r_2 \quad s=r_1-r_3 \ , \\
& c^{(3)}(r;s) = c^{(3)}(s,r) = c^{(3)}(-r;s-r) \ . \label{properties_inv_trans}
\end{split}\eeq
Defining the double fourier transform of $c^{(3)}$ as:
\beq
c^{(3)}(k,q) = \int_{r,s} e^{-i k r} e^{-i q s} c^{(3)}(r,s) \ , \label{fourier_2variables}
\eeq
we obtain two invariance principles:
\beq\label{c3inv}
c^{(3)}(k,q) = c^{(3)}(q,k) = c^{(3)}(-k-q,q) \ .
\eeq
Performing a Fourier transformation on our equation we get
\beq\begin{split}
& \tilde{c}(k) =  \FF \left( \tilde{h} - \ln [ 1 + \tilde{h} ] \right)(k)  \\
& + \frac{\r^4}2 \int_q c^{(3)}(-k,q) c^{(3)}(k,-q) \tilde{h}(q) 
\tilde{h}(k-q)  \\
& + \frac{\r^2}{2} \int_{q}  c^{(3)}(q,k-q) \tilde{h}(q) \tilde{h}(k-q) \\
& + \frac{\r^2}{2} \int_{q} c^{(3)}(-q,-k+q) \tilde{h}(q) \tilde{h}(k-q) \ , 
\end{split}\eeq
which using the invariances in Eq.~(\ref{c3inv}) is simplified to:
\beq \begin{split}
& \tilde{c}(k) =  \FF \left( \tilde{h} - \ln [ 1 + \tilde{h} ] \right)(k) \\
& + \frac{1}{2} \int_{q} \left(\left[ 1 + \r^2  c^{(3)}(q,k-q) \right]^2 -1 \right) \tilde{h}(q) \tilde{h}(k-q) \ .
\end{split}\eeq
Before turning to a numerical resolution of this equation, we can again make the naive 
expansion of the HNC term in powers of $\th$, and keep only the lowest order term, 
to obtain:
\beq
\frac{f(k)}{1-f(k)} = \frac{S(k)}{2 \r} \int_{q} \r^4 \G^{(3)}_{liq}(k,-q)^2 S(q) S(k-q) f(q) f(k-q) \ .
\eeq
This recovers exactly the three-body term in the MCT kernel Eq.~(\ref{MCT_kernel}).

\section{Three-body correlations and numerical solving}
\label{sec5}

We thus have obtained a closed equation of the order-parameter, that necessitates as an input the two- and three-body direct correlation 
functions of the liquid. We already quoted that we decided here to work within the PY approximation for the two-point functions. 
It is known~\cite{hansen} that the PY approximation~\cite{PY58}, which amounts to treat the fluid in a mean-field approximation, but under the exact constraint that the pair 
correlation function should vanish for distances smaller than $1$~\cite{ES93}, is particularly efficient for hard spheres.
 Furthermore, we dispose of an analytic expression for the two-body direct correlation function 
in that approximation~\cite{We63,Th63}. However, the three-body direct correlation function still needs to be approximated. 
Computing the third-order direct correlation function is in itself a hard problem of liquid theory. The best approximation available was shown by 
numerical works~\cite{BK94} to be the 
HNC3 approximation developed by Attard~\cite{At90}. However, this approximation is very computationally demanding, and because our purpose here
is merely to demonstrate the 
importance of higher-order terms in the expansion in powers of the order parameter, we do not aim at quantitative efficiency, and wish to 
find a simpler approximation scheme.

\subsection{Denton \& Ashcroft approximation}

A good compromise between simplicity and efficiency for evaluating the third-order direct correlation function~\cite{BK93} 
is the Denton-Ashcroft approximation~\cite{DA89}. This approximation gives an analytic form that necessitates as an input only the second-order
direct correlation function, for which we can use the PY result. Within their approximation, $c^{(3)}$ is given by:
\beq \begin{split}
c^{(3)}_{DA}(k,q) = & \frac 1{c^{(2)}(0)} \left[ c^{(2)}(k) \partial_\rho c^{(2)}(q) + c^{(2)}(q) \partial_{\rho} c^{(2)}
(k) \right]  \\
& - \frac{\partial_\rho c^{(2)}(0)}{\left( c^{(2)}(0) \right)^2} c^{(2)}(k) c^{(2)}(q) \ .
\label{DA_nonsymmetrized}
\end{split} \eeq
Within this approximation, angular dependance is neglected. We can recover it by symmetrizing the expression:
\beq \begin{split}
c^{(3)}_{DAS}(k,q) = \frac 13 & \left[ c^{(3)}_{DA}(k,q) + c^{(3)}_{DA}(k,|k+q|) \right.  \\
& \left. + c^{(3)}_{DA}(q,|k+q|) \right] \ .
\end{split} \eeq

As stated before, we use as an input the Percus-Yevick direct correlation function, that reads (in units of the hard-sphere diameter):
\beq \begin{split}
c^{(2)}_{PY}(r) = & \left\{ \begin{array}{ll}
\displaystyle - a - 6 \varphi b r - \frac 12 \varphi a r^3 & r \le 1 \ , \\
\displaystyle 0 &  r > 1 \ . \end{array} \right. 
\label{c_PY}
\end{split} \eeq
where $ \varphi$ is the packing fraction defined by $\varphi = \p \r / 6$, $a=(1+2\varphi)^2/(1-\varphi)^4$ and $b=-(1+\varphi/2)^2/(1-\varphi)^4$.
The corresponding Fourier transforms, and derivatives with respect to density are simply computed analytically.
Using this approximation for $c^{(3)}$, and writing the integrals in bipolar coordinates by using the isotropy of the liquid,
we obtain the following set of equations:
\begin{widetext}
\beq
\left\{ \begin{array}{l l}
\tilde{c}(k) & \displaystyle = F(k) + H(k) \ , \\
& \\
F(r) & = \displaystyle \tilde{h}(r) - \ln \left( 1 + \tilde{h}(r) \right) \ , \\
& \\
H(k) & \displaystyle = \frac{\r S(k)}{8 \p^2 k}  \int_0^\infty du \int_{|k-u|}^{k+u} dv ~ u ~ v \tilde{h}(u) \tilde{h}(v) 
\left( \left[ 1+ \rho^2 {\hat c}^{(3)}_{DAS}(u,v;k) \right]^2 - 1 \right) \ , \\
\end{array} \right.  \label{RHNC3body} 
\eeq
\end{widetext}
where we have defined
\beq
{\hat c}^{(3)}_{DAS}(u,v;k) = \frac 13 \left[ c^{(3)}_{DA}(k,u) + c^{(3)}_{DA}(k,v) + c^{(3)}_{DA}(u,v) \right] \ .
\label{final_4}
\eeq
Equations (\ref{DA_nonsymmetrized}) -- (\ref{final_4}) now completely specify our approximation. We present in the following a preliminary numerical 
resolution in order to demonstrate the importance of the correction $H$.

\subsection{Numerical resolution methodology}

The usual calculation for HNC alone is more stable if we write the iteration procedure in terms of $\tilde{c}$ and $\tilde{\chi}=\tilde{h}-\tilde{c}$, which would give in 
our case:
\beq
\tilde{c}(r) = e^{\tilde{\chi}(r)+H(r)}-1 - \tilde{\chi}(r) \ ,
\label{HNC_3body}
\eeq
or the same equation with $H=0$ in the case of RHNC.
We first solve the RHNC equation Eq.~(\ref{RHNC3body}) with $H=0$, by using PY approximation for the two-point functions. 
Once a stable solution of the RHNC equations has been found
we introduce the three-body correction $H$, and solve by the very same Picard iterative scheme that is used for solving HNC and RHNC.
Explicitly, the resolution will give:
\begin{itemize}
\item Start from the value of $k \tilde{c}(k)$, obtained with the previous iteration, or from the old HNC value if first iteration
\item Use Eq.~(\ref{OZ2rep_fourier}) to deduce the value of $k \tilde{h}(k)$ and $k \tilde{\chi}(k)$
\item Use $k \tilde{h}(k)$ to evaluate $k H(k)$
\item Inverse Fourier transform $k \tilde{\chi}(k)$ and $k H(k)$ to obtain $r \tilde{\chi}(r)$ and $r H(r)$
\item Use it to evaluate the new $r \tilde{c}(r)$ with Eq.~(\ref{HNC_3body})
\item Fourier transform $r \tilde{c}(r)$ to obtain $k \tilde{c}(k)$
\item Mix it with the old $k \tilde{c}(k)$ to avoid rapid changes 
\item Repeat these steps until $k \tilde{c}(k)$ has converged
\end{itemize}
We used a grid of $2^{10}$ equally spaced points on a box of size $11$, and a mixing parameter $0.01$. We note that evaluation of the 
correcting term $H$ in Eq.~(\ref{RHNC3body}) has a computational cost of order of the square number of points on the grid, significantly slowing down 
the resolution of the equation, since to avoid instabilities, $\tilde{c}$ is made to evolve very slowly by the mixing procedure.

\begin{figure}[t]
\includegraphics[width=8cm]{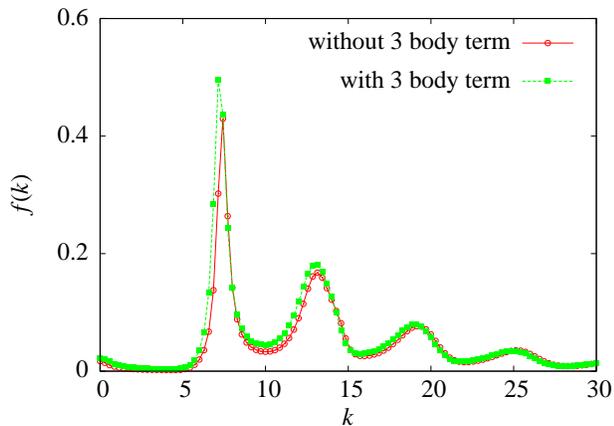}
\caption{Non-ergodicity factor as a function of the wave vector very close to the critical point $\varphi=\varphi_d$, 
for the PY + HNC calculation, and with inclusion of the three body term.}
\label{fig:compPYRHNC_3bodyRHNC}
\end{figure}

\subsection{Results and discussion}

We solved our improved equation~(\ref{RHNC3body}) for several packing fractions 
starting from $\varphi = 0.6$, and decreasing it until the non-trivial solution for
$\th$ disappeared. We found that the inclusion of the three-body terms, even when 
using a crude approximation such as the symmetrized Denton \& Ashcroft 
approximation, leads to a strong shift of the dynamical transition point from
$\varphi_d = 0.591897$ to $\varphi_d = 0.555860$, which is closer to the 
predicted MCT transition but farther from the numerically estimated transition.
It is hard to guess what would be the evolution of $\varphi_d$ when additional corrections are added.

The resulting non-ergodicity factors at the transition are 
depicted in 
Fig.~\ref{fig:compPYRHNC_3bodyRHNC}. 
The inclusion of the three body term has 
significantly enhanced the first peak of $f(k)$ upon inclusion of our correction.
These preliminary results must be treated with caution, because it was found~\cite{FJPUZ12b} that 
the numerical solving of the RHNC equation is sensitive to the discretization used, and 
very large grids with a very large number of points must be used in order to obtain
stable results. This situation is expected to be the same with the presence of the 3 body 
term, but the $\OO(N^2)$ scaling of the numerical resolution in our case prevented us
from performing a stability analysis. The qualitative picture is nevertheless not expected 
to be modified by these considerations.

\begin{figure}[t]
\includegraphics[width=8cm]{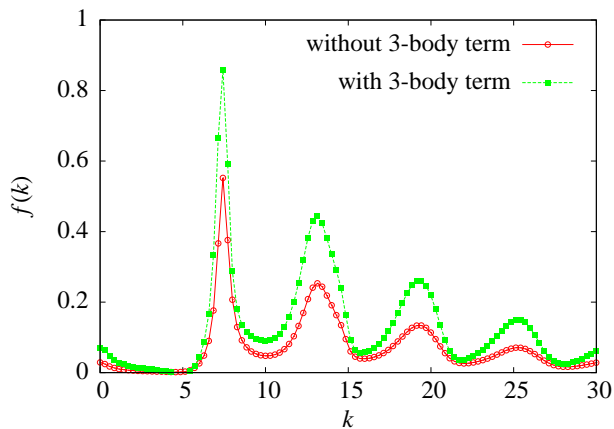}
\caption{Non-ergodicity factor as a function of the wave vector at fixed packing fraction $0.6$, for the PY + RHNC calculation, and with inclusion of the three body term.
}
\label{fig:compPYRHNC_3bodyRHNC_fixedphi}
\end{figure}

In Fig.~\ref{fig:compPYRHNC_3bodyRHNC_fixedphi} we show the results inside the glass phase, at a fixed packing fraction
$\varphi = 0.6$, with and without inclusion of the three-body term.
In this case the effect of including the 3-body term is much bigger and goes in the right direction of increasing $f(k)$ at small
$k$ (although still not enough at very small $k$ below the first peak).

\section{Higher orders}
\label{next_orders}

Our calculation provides the {\it exact} free-energy and correlation function at order $\tilde{h}^2$. This feature allows us to put into
correspondence the MCT kernel, which is also $\OO(f^2)$, with the replica result. We have seen that the MCT kernel in 
Eq.~(\ref{MCT_kernel}) contains exactly the three-body contribution we obtain with replica theory, which is already a quite surprising
result, given the differences that exist between the two approaches. However, it shows that the two-body terms contained in 
the MCT kernel are thus impossible to obtain within a static framework. The peculiar wave-vector dependance of these terms arise 
from the calculation of forces, inherent in dynamical theories, but absent in static ones. One could however wonder whether these
two-body terms could arise in a dynamical calculation because of the factorization of a dynamical four-point vertex function, that would 
be forced to be expressed as a $\OO(f^2)$ function. Indeed, the main approximation involved in Mode-Coupling theories is the 
factorization of a four-point function, and since MCT breaks down at high dimensions, it is possible that in the process, ``glassy" 
correlations are factorized along with ``liquid" ones, forcefully introducing these new $\OO(f^2)$ terms. This is nevertheless highly
speculative, and no satisfying dynamical theory exists yet, that would be able to investigate these considerations (note however two
formulations in \cite{Sz07} and \cite{Ma11} that may have this potential). 

\begin{figure}[ht]
\begin{center}
\fcolorbox{white}{white}{
  \begin{picture}(164,130) (0,0)
    \SetWidth{1.0}
    \SetColor{Black}
    \Oval(17,65)(64,16)(0)
    \Oval(129,65)(64,16)(0)
    \Vertex(33,49){6}
    \Vertex(113,49){6}
    \Photon(33,49)(113,49){3.5}{6}
    \Vertex(33,81){6}
    \Vertex(113,81){6}
    \Photon(33,81)(113,81){3.5}{6}
    \Vertex(28,113){6}
    \Vertex(118,113){6}
    \Photon(28,113)(117,113){3.5}{7}
    \Vertex(28,17){6}
    \Vertex(119,17){6}
    \Photon(28,17)(117,17){3.5}{7}
    \Text(13,65)[lb]{\Large{\Black{$a$}}}
    \Text(126,65)[lb]{\Large{\Black{$b$}}}
  \end{picture}
}
\end{center}
\caption{Diagrams that contribute to the free-energy at order $\tilde{h}^4$. A wiggly line joining two replica indices $a$ and $b$ 
is a $h_{ab}$, with $a \ne b$ function, a black dot attached to a zone with replica index $a$ is an integration point weighted by a 
density factor $\r_a$.}
\label{fig:order_4}
\end{figure}
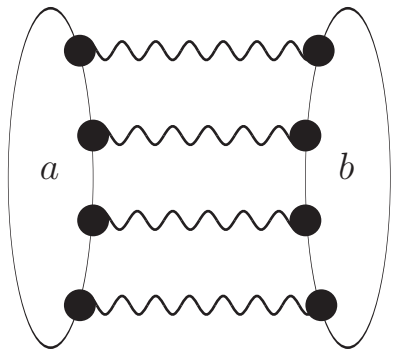

An interesting feature of our calculation is that, once the theory is set up, we can already uncover the next terms with a 
diagrammatical visualization of the expansion. To go further, we can now wonder what is the term $\OO(\tilde{h}^4)$ in the 
free-energy (that would correspond to a $\OO(f^3)$ term for the correlation function). The requirement that we must have 2PI 
diagrams is quite strong, and it is easy to convince one-self that the only possible diagram that we can construct is shown in 
Figure~\ref{fig:order_4}. The four-point functions that are connected by $\tilde{h}$ lines are made of 1PI diagrams, and we will
thus obtain a new contribution to the free-energy that contains the 4-point vertex functions of the liquid:
\beq \begin{split}
\text{const} & \times \int_{r_1,\ldots,r_8} \G^{(4)}_{liq}(r_1,\ldots,r_7) \G^{(4)}_{liq}(r_2,\ldots,r_8) \\
& \times \tilde{h}(r_1,r_2) \cdots \tilde{h}(r_7,r_8) \ .
\end{split} \eeq
Interestingly, we don't even have to work out the precise diagrammatics behind this procedure, since RHNC gives a contribution
in every diagram, and at all orders in $\tilde{h}$, so that we can use RHNC to fix the prefactor of each diagram we compute.

Computing the $n$-th order direct correlation function is still a difficult problem, but with approximations such as the Denton-Ashcroft
approximation, that are issued from density functional theory, we can deduce from the approximation for $c^{(3)}$ the corresponding
approximation for $c^{(n)}$ by successive differentiation with respect to density. The last difficulty is that an $\OO(f^n)$ kernel will
require the numerical evaluation of an $n$-dimensional integral, that have a computational cost of order $\OO(N^n)$.

\begin{figure}[ht]
\fcolorbox{white}{white}{
  \begin{picture}(162,130) (0,0)
    \SetWidth{1.0}
    \SetColor{Black}
    \Oval(17,65)(64,16)(0)
    \Oval(130,65)(64,16)(0)
    \Vertex(28,112){6}
    \Vertex(118,112){6}
    \Photon(28,112)(117,112){3.5}{7}
    \Text(13,67)[lb]{\Large{\Black{$a$}}}
    \Text(127,67)[lb]{\Large{\Black{$b$}}}
    \Vertex(34,70){6}
    \Vertex(112,70){6}
    \Photon(34,70)(112,70){3.5}{6}
    \Vertex(32,90){6}
    \Vertex(114,90){6}
    \Photon(32,90)(114,90){3.5}{6}
    \Vertex(32,48){6}
    \Vertex(114,48){6}
    \Photon(32,48)(114,48){3.5}{6}
    \Vertex(28,26){6}
    \Vertex(118,26){6}
    \Photon(28,26)(116,26){3.5}{7}
  \end{picture}
}
\fcolorbox{white}{white}{
  \begin{picture}(120,131) (0,0)
    \SetWidth{0.5}
    \SetColor{Black}
    \Text(84,104)[lb]{\Large{\Black{$b$}}}
    \SetWidth{1.0}
    \Vertex(90,82){7}
    \Text(12,63)[lb]{\Large{\Black{$a$}}}
    \Oval(16,63)(62,15)(0)
    \Oval(85,104)(32,13)(40)
    \Oval(82,28)(32,13)(-40)
    \Vertex(27,23){7}
    \Vertex(72,37){7}
    \Vertex(31,79){7}
    \Vertex(31,50){7}
    \Vertex(88,49){7}
    \Vertex(76,95){7}
    \Vertex(65,110){7}
    \Vertex(61,19){7}
    \Vertex(27,105){7}
    \Photon(26,106)(60,111){4}{3}
    \Photon(30,77)(74,96){4}{4}
    \Photon(31,51)(72,40){4}{4}
    \Photon(27,24)(59,20){4}{3}
    \Photon(89,52)(89,84){4}{3}
    \Text(82,26)[lb]{\Large{\Black{$c$}}}
  \end{picture}
}
\caption{Diagrams that contribute to the free-energy at order $\tilde{h}^5$. A wiggly line joining two replica indices $a$ and $b$ 
is a $h_{ab}$, with $a \ne b$ function, a black dot attached to a zone with replica index $a$ is an integration point weighted by a 
density factor $\r_a$.}
\label{fig:order_5}
\end{figure}
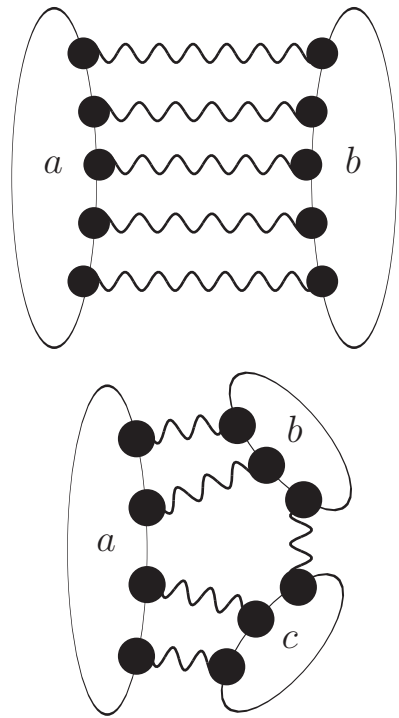

In order to visualize the increasing difficulty of going to the next orders, we show in Figure~\ref{fig:order_5} the possible diagrams at 
order $\tilde{h}^5$. We see that new, intricate terms arise, and that in general, the $\OO(\tilde{h}^n)$ contribution will contain all 
vertex functions of the liquid of orders ranging from $3$ to $n$.

\section{Conclusion and discussion}
\label{conclusion}

Our analysis shows that the 2PI corrections to the RHNC free energy are quantitatively
relevant at the dynamical transition, and must be properly taken into account in order to
obtain an accurate static description of the transition. It seems then that only by doing this 
properly we will be able to make a clear connection between dynamical and static theories of the 
dynamical transition of glasses. 
In this respect our results are the following:
\begin{itemize}
\item At the level of the RHNC approximation, the equation for $f(k)$ can be developed
in powers of $f(k)$. The result is Eq.~(\ref{eq_f_HNC}), which corresponds to the Mode-Coupling
equation with a kernel equal to 1. The latter is also the result of a zero-th order field theory calculation as
reported in~\cite{ABL06}.
\item The first correction to the RHNC approximation provides an additional contribution
to the kernel which happens to correspond exactly to the three-body term of the Mode-Coupling kernel.
\item The next corrections will give terms proportional to $f^3$ in the right hand side of Eq.~(\ref{eq_f_HNC}),
hence no additional contributions to the Mode-Coupling kernel can be generated by these terms.
We are forced to conclude that there is no way of generating the terms of the Mode-Coupling kernel
proportional to $c(k)$ by means of a static computation.
\item We have shown the way to compute higher-order terms, although we expect the numerical resolution of the corresponding
approximations to be hard.
\end{itemize}
It would be therefore very important to perform a similar calculation (namely, a systematic expansion in powers of $f(k)$) 
also on the dynamical side. This would allow for a systematic comparison of the results. 
One would then obtain a proper theory for the ergodicity breaking that 
occurs at $\r_d$, free of the ambiguities of MCT, and systematically improvable. Work 
has been done in this direction in the last years~\cite{ABL06,KK07,ABL09,JW11}, but 
the situation is still unsatisfactory. 

\begin{figure}[ht]
\includegraphics[width=8cm]{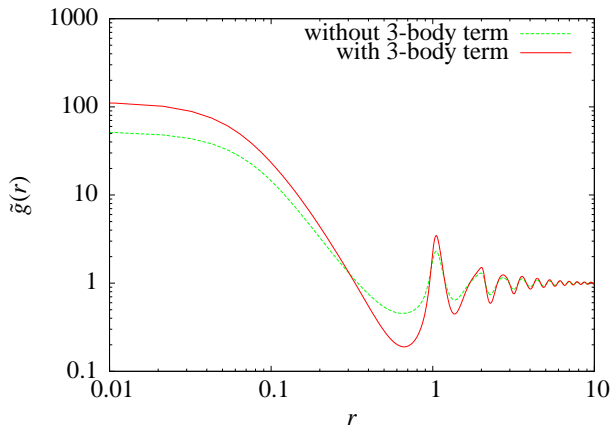}
\caption{Inter-replica pair correlation function $\tilde{g}(r)$ at packing fraction $\varphi=0.6$, with and without three-body correction.}
\label{fig:gtilde_r}
\end{figure}

On the static side, the first task to be performed is to reconcile the small cage expansion
and the 2PI approach that we use in this work in order to have a unified static theory.
The way to do this is indicated by the shape of the pair correlation function between 
two different replicas at the transition, depicted in Fig. \ref{fig:gtilde_r}.
We see that $\tilde{g}$ develops a strong peak at $r=0$, that dominates the rest of its
features. This peak simply reflects the fact that, in the glass phase, two replicas tend
to be very close from each other. In the small cage expansion, this idea is reflected in
the introduction of a cage size parameter $A$, which serves as a expansion parameter.
In our case a direct introduction of the cage size $A$ would be more difficult, because it 
would amount to parametrize the pair correlation function by the cage size, which is difficult
to do without specifying a given shape of cages. In the high dimensional limit, the 
cage can safely be approximated by a Gaussian as far as the free energy is concerned~\cite{KPZ12}, 
but this is not expected to hold in our low dimensional case~\cite{CIPZ12}.

It is interesting to note that we have observed, as in Fig.~\ref{fig:gtilde_r} that the inclusion of the three-body term systematically
increases the spatial separation between the first, glassy, peak of $\tilde{g}(r)$ and the subsequent, liquid peaks.
A clear separation between these contributions allows for an unambiguous definition of the molecules
introduced in order to perform small-cage expansions \cite{MP99b,PZ10}, and is clearly a sign that reconciling
RHNC with small-cage expansions may not be out of reach.

The large contribution coming from the peak at $r=0$ in $\tilde{g}$ shows 
that the diagrams that contribute the most to the free-energy are the most connected 
ones~\cite{PZ10}. It should be possible to put these diagrams in correspondence with the diagrams
re-summed in the small cage expansions of~\cite{MP99b} and~\cite{PZ10} in order to 
make progress. Our work has set up the tools necessary to perform such resummations 
and we believe this is the natural line of work to follow in the future.

\acknowledgments
H.J. PhD work is funded by a CFM -- JP Aguilar grant. We acknowledge discussions
with Alexei Andreanov, Jean-Louis Barrat, Ludovic Berthier, Giulio Biroli, Patrick Charbonneau,
Daniele Coslovich, Silvio Franz, Atsushi Ikeda, Giorgio Parisi, Grzegorz Szamel, 
Pierfrancesco Urbani and Fr\'ed\'eric van Wijland.

\bibliographystyle{plain}

\end{document}